\documentclass[11pt]{article}
\usepackage{jheppub}
\usepackage{graphicx,array,dcolumn}
\usepackage{calc,tabularx, epsfig,mathrsfs}
\usepackage{hyperref}

\usepackage{amsmath,verbatim,enumerate}
\usepackage{amssymb}
\usepackage{wasysym}
\usepackage{xcolor}
\usepackage{multirow}
\usepackage{xspace}
\usepackage{slashed}
\allowdisplaybreaks[1]
\newlength{\figurewidth}
\newcommand{\beq}{\begin{equation}}
\newcommand{\eeq}{\end{equation}}
\newcommand{\bea}{\begin{eqnarray}}
\newcommand{\eea}{\end{eqnarray}}
\newcommand{\ba}{\begin{array}}
\newcommand{\ea}{\end{array}}

\newcommand{\mn}{{\mu\nu}}

\newcommand{\pt}{\partial}

\newcommand{\al}{\alpha}

\newcommand{\ep}{\epsilon}
\newcommand{\ta}{\theta}
\newcommand{\lam}{\lambda}
\newcommand{\Lam}{\Lambda}
\newcommand{\G}{\Gamma}

\newcommand{\de}{\delta}
\newcommand{\D}{\Delta}
\newcommand{\OM}{\Omega}
\newcommand{\om}{\omega}
\newcommand{\sg}{\sigma}

\makeatother
\title{On Gauss-bonnet gravity and boundary conditions in Lorentzian path-integral quantization}
\author[a]{Gaurav Narain}
\affiliation[a]{
Center for Gravitational Physics, Department of Space Science, 
Beihang University, Beijing 100191, China.
}
\emailAdd{gaunarain@gmail.com}

\abstract{
Recently there has been a surge of interest in studying Lorentzian 
quantum cosmology using Picard-Lefschetz methods. 
The present paper aims to explore the Lorentzian path-integral of Gauss-Bonnet gravity in 
four spacetime dimensions with metric as the field variable.
We employ mini-superspace approximation and study the 
variational problem exploring different boundary conditions. 
It is seen that for mixed boundary conditions 
non-trivial effects arise from Gauss-Bonnet sector of gravity
leading to additional saddle points for lapse in some case. 
As an application of this we consider the 
No-boundary proposal of the Universe with two different settings
of boundary conditions, and compute the transition amplitude 
using Picard-Lefschetz formalism. 
In first case the transition amplitude is a superposition 
of a Lorentzian and a Euclidean geometrical configuration 
leading to interference incorporating 
non-perturbative effects coming from Gauss-Bonnet sector of gravity.
In the second case involving complex 
initial momentum we note that the transition amplitude is an analogue of
Hartle-Hawking wave-function with non-perturbative correction 
coming from Gauss-Bonnet sector of gravity. 
}

\begin{document}
\maketitle

\section{Introduction}
\label{intro}

General relativity (GR) is the first theory to unite gravity with relativity using the notion 
of curved spacetime. It has been hugely successful at offering explanations 
for a variety of physical phenomenon ranging from astrophysical to cosmological 
scales. Despite its huge successes as a classical theory, the theory suffers 
from lack of completion at small scales as also is demonstrated by singularity 
theorems. When the successful framework of quantum field theory (QFT) is applied to GR, 
then the QFT of GR is seen to be non-renormalizable and the resulting theory 
lacks predictivity 
\cite{tHooft:1974toh,Deser:1974nb,Deser:1974cz,Deser:1974cy,Goroff:1985sz,Goroff:1985th,vandeVen:1991gw}.
These results although perturbative indicate the lack of compatibility of the two 
theories at small scales thereby suggesting that either GR or QFT or 
both should suitably be modified at short distances to have a well-defined 
ultraviolet complete theory of gravity. 

Experimental observations at large distances fail to agree with the predictions of 
GR coupled with standard model. This although doesn't imply failure of GR, but 
it does hint that the coupled system of standard model and GR is missing something. 
Over the years researchers have tried to address this issue in different ways: (1) modify 
the standard model by incorporating dark-matter and dark-energy while keeping GR to be 
unmodified, (2) modify gravitational dynamics at large distances keeping the 
standard model unmodified or (3) modify both GR and standard model. In the case 
(2) and (3) one goes beyond GR. 

Numerous model has been proposed over the years to explain 
such deviations at both ends of energy scales. 
For example at ultra short length scales, motivated by lack of renormalizabilty of GR
(which has only two time derivatives of the metric field) 
proposals have been made to incorporate higher-time derivatives of the metric field.
Such modifications of GR are collectively referred to as higher-derivative theories of 
gravity. It has been noticed that incorporating higher-derivatives although 
addresses issues of renormalizability in four spacetime dimensions 
but the theory lacks unitarity \cite{Stelle:1976gc,Salam:1978fd,Julve:1978xn}. 
Some efforts have been made to tackle this unwanted problem in
the perturbative framework 
\cite{Narain:2011gs,Narain:2012nf,Narain:2017tvp,Narain:2016sgk},
in asymptotic safety approach 
\cite{Codello:2006in,Niedermaier:2009zz}
and `\textit{Agravity}' \cite{Salvio:2014soa}. However, one also notices an important fact 
that equation of motion in such UV modified theories have higher-time derivatives 
(more than two) of metric field. One wonders at the presence of such 
higher-time derivatives which although addresses renormalizability 
but also are responsible for lack of tree-level unitarity in theory. 

Overtime a need arose of having a modification of GR which consist of 
higher-derivatives of the metric field, but when contributions from all such terms are summed over then
the highest order of time derivative is only two. 
The Gauss-Bonnet (GB) gravity in four spacetime dimensions is one such 
simple modification of the GR, which satisfies exactly this requirement.
Here the dynamical evolution equations of field remains unaffected. 
In four spacetime dimension GB sector of gravity action is also topological
and doesn't play any role in dynamical evolution of spacetime metric.
However, they play a key role in path-integral quantization 
of gravity where it is used to classify topologies
and has an important role to play at boundaries.
The bulk Gauss-Bonnet gravity action is following
\bea
\label{eq:act}
S = \frac{1}{16\pi G} \int {\rm d}^Dx \sqrt{-g}
\biggl[
-2\Lam + R + \al
\biggl( R_{\mu\nu\rho\sg} R^{\mu\nu\rho\sg} - 4 R_\mn R^\mn + R^2 \biggr)
\biggr] \, , 
\eea
where $G$ is the Newton's gravitational constant, 
$\Lam$ is the cosmological constant term,  
$\al$ is the Gauss-Bonnet coupling and $D$ is spacetime dimensionality. 
The mass dimensions of various couplings are: 
$[G] = M^{2-D}$, $[\Lam] = M^2$ and $[\al] = M^{-2}$. 
This action falls in the class of Lovelock-Lanczos gravity theories 
\cite{Lovelock:1971yv,Lovelock:1972vz,Lanczos:1938sf}, and
are a special class of higher-derivative gravity 
where equation of motion for the metric field remains second order in time.

Motivated by the above properties of GB gravity, it is worth asking how the 
path-integral of the metric field behaves when its gravitational action is given by 
eq. (\ref{eq:act})? My interest in this paper is to study such a path-integral 
in a slightly simpler setting: the context of quantum cosmology where the 
issues regarding various kind of boundary conditions are easier to investigate 
\cite{York:1986lje,Brown:1992bq,Krishnan:2016mcj,Witten:2018lgb,Krishnan:2017bte}.
We start by considering a generic metric respecting 
spatial homogeneity and isotropicity in $D$ spacetime 
dimensions. It is the FLRW metric in arbitrary spacetime 
dimension with dimensionality $D$. In polar
co-ordinates $\{t_p, r, \ta, \cdots \}$ the FLRW metric 
can be expressed as
\beq
\label{eq:frwmet}
{\rm d}s^2 = - N_p^2(t_p) {\rm d} t_p^2 
+ a^2(t_p) \left[
\frac{{\rm d}r^2}{1-kr^2} + r^2 {\rm d} \OM_{D-2}^2
\right] \, .
\eeq
It consists of two unknown time-dependent functions: lapse $N_p(t_p)$
and scale-factor $a(t_p)$. 
Here $k=(0, \pm 1)$ is the curvature, and ${\rm d}\OM_{D-2}$ is the 
metric corresponding to unit sphere in $D-2$ spatial dimensions. 
This is the mini-superspace approximation of the metric.
This is a gross simplification of the original gravitational theory 
in a sense as we no longer have gravitational waves.
However, we do retain the diffeomorphism invariance of the time 
co-ordinate $t_p$ and the dynamical scale-factor $a(t_p)$. 
This simple setting is enough to investigate the effects the 
GB-modification of GR in the gravitational path-integral. 

The Feynman path-integral for the reduced theory can be written as
\beq
\label{eq:Gform_sch_fpt}
G[{\rm bd}_0, {\rm bd}_1]
= \int_{{\rm bd}_0}^{{\rm bd}_1} {\cal D} N_p {\cal D} \pi {\cal D} a(t_p) {\cal D} p {\cal D} {\cal C} {\cal D} \bar{P}
\exp \biggl[
\frac{i}{\hbar} \int_0^1 {\rm d} t_p \left(
N^\prime_p \pi + a^\prime p +{\cal C}^\prime \bar{P} - N_p H \right)
\biggr] \, ,
\eeq
where beside the scale-factor $a(t_p)$, lapse $N_p$ and fermionic ghost
${\cal C}$ we also have their corresponding conjugate momenta given by
$p$, $\pi$ and $\bar{P}$ respectively. Here $({}^\prime)$ denotes derivative 
with respect to $t_p$. The original path-integral measure 
becomes a measure over all the variables. The time $t_p$ co-ordinate can be 
chosen to range from $0\leq t_p \leq 1$ without compromising on 
generality. Here ${\rm bd}_0$ and ${\rm bd}_1$ refers to field configuration at 
initial ($t_p=0$) and final ($t_p=1$) boundary respectively. 
The Hamiltonian constraint $H$ consists of two parts 
\beq
\label{eq:Htwo}
H = H_{GB}[a, p] + H_{\rm gh} [N, \pi, {\cal C}, \bar{P}] \, ,
\eeq
the Hamiltonian corresponding to Gauss-Bonnet gravity action 
is denoted by $H_{GB}$ and the Batalin-Fradkin-Vilkovisky (BFV) \cite{Batalin:1977pb}
ghost Hamiltonian is denoted by $H_{\rm gh}$
\footnote{ 
The BFV ghost is an extension of the usual Fadeev-Popov ghost which is 
based on BRST symmetry. In usual gauge theories the 
constraint algebra forms a Lie algebra, while the constraint algebra 
doesn't closes in case of diffeomorphism invariant gravitational theories. 
For this reason one needs BFV quantization process. In the case of 
mini-superspace approximation however, we have only one constraint 
$H$. In this approximation therefore the algebra trivially closes leaving the 
distinction between two quantization process irrelevant. 
Nevertheless BFV quantization is still preferable.}.
The mini-superspace approximation still retains diffeomorphism invariance 
which show up as a time reparametrization symmetry. This invariance 
can be broken by fixing the proper-time gauge $N^\prime_p=0$. 
For more elaborate discussion on BFV quantization process and ghost
see \cite{Teitelboim:1981ua,Teitelboim:1983fk,Halliwell:1988wc}. 

In the mini-superspace approximation most of the path-integral in 
eq. (\ref{eq:Gform_sch_fpt}) can be performed analytically leaving behind 
the following path-integral 
\beq
\label{eq:Gform_sch}
G[{\rm bd}_0, {\rm bd}_1]
= \int_{0^+}^{\infty} {\rm d} N_p
\int_{{\rm bd}_0}^{{\rm bd}_1} {\cal D} a(t_p) \,\,
e^{i S[a, N_p]/\hbar} \, .
\eeq
This residual path-integral is easy to interpret. The path-integral 
$\int {\cal D} a(t_p) \,\, e^{i S[a, N_p]/\hbar}$ represent the quantum-mechanical 
transitional amplitude for the Universe to evolve from one 
configuration to another in proper time $N_p$. The integration 
over lapse-function $N_p$ indicates that one has to consider paths 
of every proper duration $0<N_p<\infty$. Such a choice leads to 
causal ordering of the two field configuration ${\rm bd}_0$ and ${\rm bd}_1$ 
as shown in \cite{Teitelboim:1983fh}, where $a_0<a_1$ will imply expanding 
Universe while $a_0>a_1$ will imply contracting Universe. 
In this paper we are interested in studying this residual path-integral 
for the case of Gauss-Bonnet gravity where we expect that the choice of 
boundary configurations may or may-not give rise to non-trivial features 
coming from the Gauss-Bonnet term in the gravitational action.

Ones task is now reduced to the study of functional integral in eq. (\ref{eq:Gform_sch}).
In general the standard methodology to deal with flat spacetime 
Lorentzian functional integrals of non-gravitational QFT 
is to Wick rotate the Lorentzian time co-ordinate
and go to Euclideanised time, which for example in the current case 
will mean $t_p \to i \tau_p$, where $\tau_p$ is the Euclideanised time. 
Such a rotation of time co-ordinate analytically transforms the flat spacetime 
Lorentzian path-integral to a Euclidean path-integral with 
an exponentially suppressed weight factor, which is convergent and well-defined.

In case of gravity the situation is not so straightforward. One can in principle aim to 
directly study the following Euclidean gravitational path-integral 
\beq
\label{eq:EucQGAct}
G[{\rm bd}_0, {\rm bd}_1]
= \int_{{\rm bd}_0}^{{\rm bd}_1} {\cal D} g_\mn \exp \left(-I[g_\mn] \right) \, .
\eeq
Here $g_\mn$ is the metric whose corresponding Euclidean 
action $I[g_\mn]$ appears in the exponential. 
The motivation to study directly such an Euclidean gravitational path-integral stems 
from fact that similar but non-gravitational Euclidean path-integral 
arises in flat spacetime QFT which are obtained by analytic 
continuation of a meaningful time co-ordinate. In case of gravity 
one is skeptical about the relation between the 
Euclidean and the corresponding Lorentzian path-integral, in a sense 
whether the two can be analytically related in some way. 
This is because it is not always possible to have a meaningful time co-ordinate
in generic curved spacetime.
Moreover, the Euclidean gravitational path-integral given in eq. (\ref{eq:EucQGAct}) 
suffers from the conformal factor problem, where the path-integral over the 
scale-factor is unbounded from below \cite{Gibbons:1978ac}. This implies that the Euclidean 
gravitational path-integral is not convergent and is ill-defined. This is unlike 
the situation in flat spacetime non-gravitational QFT where the Euclidean 
path-integral of the corresponding Lorentzian path-integral is convergent 
and well-defined
\footnote{
As a side remark it should be emphasised that flat spacetime has a meaningful time 
co-ordinate and enjoys the properties of 
global symmetries to cast the Lorentz group in to 
a compact rotation group under a transformation of the time co-ordinate.
Such a beauty is not present in a generic curved spacetime. 
This implies that the standard methodology of 
Wick-rotation used for defining sensible quantum field theory (QFT) 
on flat spacetime is difficult to generalise reliably in a generic 
curved spacetime where `time' is just a parameter. 
The Feynman $+i\ep$-prescription in flat spacetime QFT
is a systematic way to choose a convergent integration contour
for an otherwise highly oscillatory integral. 
It naturally implements causality in path-integral 
in a systematic manner by requiring that the 
Euclideanised version of two-point function must satisfy Osterwalder-Schrader positivity.
Such benefits exist only in flat spacetime and don't get
automatically inherited to generic Lorentzian curved spacetime. The situation gets even 
more cumbersome when spacetime becomes dynamical due to gravity
and/or gravitational field is also quantized. Some attempts 
to incorporate Wick-rotation sensibly in curved spacetime have been
made in \cite{Candelas:1977tt,Visser:2017atf,Baldazzi:2019kim,Baldazzi:2018mtl}.
However, more work needs to be done for it to mature.}.

Picard-Lefschetz theory offers a systematic methodology to carefully handle 
oscillatory path-integrals like the one in eq. (\ref{eq:Gform_sch}). It is an extension of the 
Wick-rotation prescription to define a convergent 
functional integral on a generic curved spacetime. 
In this framework one uniquely finds contours 
in the complexified plane along which the integrand is well-behaved.
By definition the oscillatory integral along the original integration contour
becomes well-behaved and non-oscillatory along the new contour. 
Such contour-lines are termed \textit{Lefschetz thimbles}. 
This framework is based on complex analysis has been recently used 
in context of Lorentzian quantum cosmology 
\cite{Feldbrugge:2017kzv,Feldbrugge:2017fcc,Feldbrugge:2017mbc},
where the authors studied gravitational path-integral  
in the mini-superspace approximation for Einstein-Hilbert gravity
\footnote{
It should be mentioned that usage of complex analysis was also made in past to study Euclidean 
gravitational path-integrals in eq. (\ref{eq:EucQGAct}) which are know to suffer 
from conformal factor problem \cite{Hawking:1981gb,Hartle:1983ai}. 
In the context of Euclidean quantum cosmology usage of complex analysis 
was made to explore issues regarding initial conditions: 
\textit{tunnelling} proposal \cite{Vilenkin:1982de,Vilenkin:1983xq,Vilenkin:1984wp} 
and 
\textit{no-boundary} proposal \cite{Hawking:1981gb,Hartle:1983ai,Hawking:1983hj}.
}.

Once it is possible to have a well-defined convergent Lorentzian path-integral using 
Picard-Lefschetz theory, one can then explore the various choices
of allowed boundary conditions. In the context of 
Euclidean quantum gravity, whose path-integral suffers from 
conformal factor problem \cite{Gibbons:1977zz,Gibbons:1978ac}, it was realised 
that a sensible choice of initial conditions and integration contour 
leads to a well-defined convergent path-integral \cite{Halliwell:1988ik,Halliwell:1989dy,Halliwell:1990qr}.
This has motivated people to follow the same footsteps to study boundary condition 
choices in the context of Lorentzian gravitational path-integral which 
become well-defined using the framework of Picard-Lefschetz
\cite{Feldbrugge:2017kzv,Feldbrugge:2017fcc,Feldbrugge:2017mbc}. 

Motivated by these ideas we set to investigate the gravitational path-integral 
for the Gauss-Bonnet gravity in the mini-superspace approximation
using the technology of Picard-Lefschetz theory. 
We start by varying the action with respect to 
field and study the nature of surface terms. 
We explore three different choice of boundary conditions: 
Dirichlet (D), Neumann (N) and Mixed (M) boundary conditions (BC). 
In each case the surface-terms are either zero or contribute same as in GR,
except in the case of mixed-boundary conditions (MBC) where the surface 
terms gets an additional non-trivial contribution from Gauss-Bonnet sector of 
gravitational action. This has interesting consequences in the followup 
study of the path-integral. Mixed boundary conditions (MBC) are interesting 
and have also been previously explored in the context of 
Einstein-Hilbert gravity \cite{DiTucci:2019dji,DiTucci:2019bui} in relation with 
no-boundary proposal of the Universe. We explore MBC 
in the context of gravitational path-integral of Gauss-Bonnet gravity, 
and find some non-trivial contribution coming from Gauss-Bonnet sector. 
As a special case we consider the no-boundary proposal 
of the Universe and find interesting features arising from the 
Gauss-Bonnet sector of gravity action. 

The outline of paper is follows: in section \ref{intro} we motivates our interest in studying this 
problem. In section \ref{minisup} we discuss the mini-superspace approximation
and compute the mini-superspace action of theory. In section 
\ref{bound_act} we discuss the action variation and study the various boundary 
conditions. In section \ref{TranProb} we consider the path-integral of gravity in 
mini-superspace approximation and start to compute the transition probability 
in saddle point approximation. Section \ref{Ninnt} studies the integration 
over lapse in complex space via Picard-Lefschetz. In section 
\ref{nbu} we study the no-boundary proposal of Universe with mixed 
boundary conditions. In section \ref{cmom} we analyse the 
Hartle-Hawking wave-function using Lorentzian path-integral and the 
corrections it receive due to Gauss-Bonnet sector of gravity. 
We finish off by presenting a conclusion and outlook in section \ref{conc}.

\section{Mini-superspace action}
\label{minisup}

The FLRW metric given in eq. (\ref{eq:frwmet}) is conformally related to flat metric and hence
its Weyl-tensor $C_{\mu\nu\rho\sg} =0$. The non-zero 
entries of the Riemann tensor are 
\cite{Deruelle:1989fj,Tangherlini:1963bw,Tangherlini:1986bw} 
\bea
\label{eq:riemann}
R_{0i0j} &=& - \left(\frac{a^{\prime\prime}}{a} - \frac{a^\prime N_p^\prime}{a N_p} \right) g_{ij} \, , 
\notag \\
R_{ijkl} &=& \left(\frac{k}{a^2} + \frac{a^{\prime2}}{N_p^2 a^2} \right)
\left(g_{ik} g_{jl} - g_{il} g_{jk} \right) \, ,
\eea
where $g_{ij}$ is the spatial part of the FLRW metric
and $({}^\prime)$ denotes derivative with respect to $t_p$.
For the Ricci-tensor the non-zero components are 
\bea
\label{eq:Ricci-ten}
R_{00} &=& - (D-1) \left(\frac{a^{\prime\prime}}{a} - \frac{a^\prime N_p^\prime}{a N_p} \right)
\, , 
\notag \\
R_{ij} &=& \left[
\frac{(D-2) (k N_p^2 + a^{\prime2})}{N_p^2 a^2}
+ \frac{a^{\prime\prime} N_p - a^\prime N_p^\prime}{a N_p^3} 
\right] g_{ij} \, ,
\eea
while the Ricci-scalar for FLRW is given by
\beq
\label{eq:Ricci0}
R = 2(D-1) \left[\frac{a^{\prime\prime} N_p - a^\prime N_p^\prime}{a N_p^3} 
+ \frac{(D-2)(k N_p^2 + a^{\prime2})}{2N_p^2 a^2} \right]
\, .
\eeq
In the case of Weyl-flat metrics one can express 
Riemann tensor in terms of Ricci-tensor and Ricci scalar.
\bea
\label{eq:Riem_exp}
R_{\mu\nu\rho\sg} = \frac{R_{\mu\rho} g_{\nu\sg} - R_{\mu\sg}g_{\nu\rho}
+ R_{\nu\sg} g_{\mu\rho} - R_{\nu\rho} g_{\mu\sg}}{D-2}
- \frac{R (g_{\mu\rho} g_{\nu\sg} - g_{\mu\sg} g_{\nu\rho})}{(D-1)(D-2)} \,.
\eea
Due to this identity we have
\beq
\label{eq:Reim2_exp}
R_{\mu\nu\rho\sg} R^{\mu\nu\rho\sg}
= \frac{4}{D-2} R_\mn R^\mn - \frac{2 R^2}{(D-1)(D-2)} \, .
\eeq
It allow us to simplify our Gauss-Bonnet gravity action for Weyl-flat metrics. 
\bea
\label{eq:actGB}
\int {\rm d}^Dx \sqrt{-g} && \left(
R_{\mu\nu\rho\sg} R^{\mu\nu\rho\sg}  - 4 R_\mn R^\mn + R^2
\right)
\notag \\
&&
= \frac{D-3}{D-2} \int {\rm d}^Dx \sqrt{-g} \left(
- R_\mn R^\mn + \frac{D R^2}{D-1}
\right) \, .
\eea
On plugging the FLRW metric of eq. (\ref{eq:frwmet}) in the action 
in eq. (\ref{eq:act}) we get an action for $a(t_p)$ and 
$N_p(t_p)$. 
\bea
\label{eq:midSact}
&&
S = \frac{V_{D-1}}{16 \pi G} \int {\rm d}t_p
\biggl[
\frac{a^{D-3}}{N_p^2} \biggl\{
(D-1)(D-2) k N_p^3 - 2 \Lam a^2 N_p^3 - 2 (D-1) a a^\prime N_p^\prime
\notag \\
&&
+ (D-1)(D-2) a^{\prime2} N_p + 2 (D-1) N_p a a^{\prime\prime}
\biggr\}
+ (D-1)(D-2)(D-3) \al\biggl\{
\frac{a^{D-5}(D-4)}{N_p^3} 
\notag\\
&&
\times (kN_p^2 + a^{\prime2})^2 
+ 4 a^{D-4}\frac{{\rm d}}{{\rm d}t_p} 
\left(
\frac{k a^\prime}{N_p} + \frac{a^{\prime 3}}{3N_p^3}
\right)
\biggr\}
\biggr] \, ,
\eea
where $V_{D-1}$ is the volume of $D-1$ dimensional space
and is given by,
\beq
\label{eq:volDm1}
V_{D-1} = \frac{\G(1/2)}{\G(D/2)} \left(\frac{\pi}{k}\right)^{(D-1)/2} \, .
\eeq
In $D=4$ we notice that in the GB-sector terms proportional 
$\al$ either vanish or are total time-derivatives. The mini-superspace 
gravitational action becomes following in $D=4$
\beq
\label{eq:mini_sup_d4}
S = \frac{V_{3}}{16 \pi G} \int {\rm d}t_p
\biggl[
6k a N_p - 2 \Lam a^3 N_p - \frac{6 a^2 a^\prime N_p^\prime}{N_p}
+ \frac{6 a a^{\prime 2}}{N_p} + \frac{6 a^{\prime\prime} a^2}{N_p}
+ 24 \al \frac{{\rm d}}{{\rm d}t_p} 
\left(
\frac{k a^\prime}{N_p} + \frac{a^{\prime 3}}{3N_p^3}
\right)
\biggr] \, ,
\eeq
This action can be recast in to a more appealing form by a 
rescaling of lapse and scale factor. 
\beq
\label{eq:rescale}
N_p(t_p) {\rm d} t_p = \frac{N(t)}{a(t)} {\rm d} t \, ,
\hspace{5mm}
q(t) = a^2(t) \, .
\eeq
This set of transformation changes our original metric in eq. (\ref{eq:frwmet})
into following
\beq
\label{eq:frwmet_changed}
{\rm d}s^2 = - \frac{N^2}{q(t)} {\rm d} t^2 
+ q(t) \left[
\frac{{\rm d}r^2}{1-kr^2} + r^2 {\rm d} \OM_{D-2}^2
\right] \, ,
\eeq
and our action in $D=4$ given in eq. (\ref{eq:mini_sup_d4}) changes to following simple form
\bea
\label{eq:Sact_frw_simp}
S = \frac{V_3}{16 \pi G} \int {\rm d}t \biggl[
(6 k - 2\Lam q) N + \frac{3 \dot{q}^2}{2N}
+ 3q \frac{{\rm d}}{{\rm d} t} \left(\frac{\dot{q}}{N} \right)
+ 24 \al \frac{{\rm d}}{{\rm d} t} \left(
\frac{k \dot{q}}{2N} + \frac{\dot{q}^3}{24 N^3} 
\right)
\biggl] \, .
\eea
Here $(\dot{})$ here represent derivative with respect to time $t$. It is worth noting 
the GB-part of action appears as a total derivative term. It will later be seen that this part 
plays a crucial role in the action for the lapse $N$ and will result in additional saddle points.
In the path-integral this term will play crucial role as it will in some sense be 
incorporating topological corrections.

\section{Boundary action and boundary conditions}
\label{bound_act}

To find the boundary action and the relevant set of boundary conditions we start 
by varying the action in eq. (\ref{eq:Sact_frw_simp}) with respect to $q(t)$. 
From now on we work in the ADM gauge $\dot{N}=0$, which implies 
that $N(t) = N_c$ (constant). We write 
\beq
\label{eq:qfluc}
q(t) = \bar{q}(t) + \ep \de q(t)
\eeq
where $\bar{q}(t)$ satisfy the equation of motion, $\de q(t)$ is the fluctuation 
around it and $\ep$ is parameter used to keep a track of the order of fluctuation terms. 
Plugging this in eq. (\ref{eq:Sact_frw_simp}) and expanding to first order in $\ep$ we have
\beq
\label{eq:Sexp_qvar}
\de S = \frac{\ep V_3}{16 \pi G} \int_{0}^{1} {\rm d}t \biggl[
\left(-2 \Lam N_c + \frac{3 \ddot{q}}{N_c} \right) \de q
+ \frac{3}{N_c} \frac{{\rm d}}{{\rm d} t} \left(q \de \dot{q} \right)
+ 24 \al \frac{{\rm d}}{{\rm d} t} \left\{ 
\left(\frac{k}{2N_c} + \frac{\dot{q}^2}{8N_c^3} \right) \de \dot{q} \right\}
\biggl] \, .
\eeq
There will also be second order terms, but for the purpose of having a sensible 
boundary value problem for the equation of motion this is sufficient. We notice that 
there are two total time-derivative pieces in the above equation which will be 
responsible for fixing appropriate boundary conditions.
The term proportional to $\de q$ gives the equation of motion for $q$
\beq
\label{eq:dyn_q_eq}
\ddot{q} = \frac{2}{3} \Lam N_c^2 \, .
\eeq
This is easy to solve and its general solution is 
\beq
\label{eq:qsol_gen}
q(t) = \frac{\Lam N_c^2}{3} t^2 + c_1 t + c_2 \, ,
\eeq
where $c_{1,2}$ are constants and will be determined based on the boundary conditions.
The total-derivative terms in the above will result in a collection of 
boundary terms
\beq
\label{eq:Sbd}
S_{\rm bdy} = \frac{\ep V_3}{16 \pi G} \biggl[
\frac{3}{N_c} \left(q_1 \de \dot{q}_1 - q_0 \de \dot{q}_0 \right)
+ 24 \al \left\{ 
\left(\frac{k \de \dot{q}_1}{2N_c} + \frac{\dot{q}_1^2\de \dot{q}_1}{8N_c^3} \right) 
-  \left(\frac{k \de \dot{q}_0}{2N_c} + \frac{\dot{q}_0^2\de \dot{q}_0}{8N_c^3} \right)\right\}
\biggr] \, ,
\eeq
where 
\beq
\label{eq:BC_name}
q_0 = q(t=0) \, , 
\hspace{5mm}
q_1 = q(t=1) \, ,
\hspace{5mm}
\dot{q}_0 = \dot{q}(t=0) \, ,
\hspace{5mm}
\dot{q}_1 = \dot{q}(t=1) \, .
\eeq

\subsection{Neumann Boundary condition (NBC)}
\label{neumann}

If we impose Neumann boundary condition (NBC) which is fixing $\dot{q}$ at both the ends of the 
$q$-trajectory \cite{Krishnan:2016mcj,DiTucci:2019dji}. 
Then we notice that the surface term in eq. (\ref{eq:Sbd}) vanish completely. 
\beq
\label{eq:neuMa_cond}
\left. \dot{q}_{0,1} \right|_{\rm NBC} = {\rm fixed} 
\hspace{3mm} 
\Rightarrow 
\hspace{3mm}
\left. \de \dot{q}_{0,1} \right|_{\rm NBC}= 0 \, ,
\eeq
where the $|_{\rm NBC}$ refers to imposing Neumann boundary condition. However, 
it is soon realised that with this boundary condition the constant $c_{1,2}$ 
appearing in the solution to equation of motion (\ref{eq:qsol_gen}) cannot be fixed
uniquely. In particular $c_2$ is left undetermined while $c_1$ will have two different 
values. This implies that it is not a well-posed problem as it leads to inconsistencies. 
This boundary condition cannot be imposed even though the surface term 
in eq. (\ref{eq:Sbd}) vanishes entirely and one doesn't
have to incorporate any additional boundary action.

\subsection{Dirichlet Boundary condition (DBC)}
\label{dirichlet}

In this boundary condition we fix the value of $q$ at the two end points. This means 
we have 
\beq
\label{eq:diRichBC}
\left. q_{0,1} \right|_{\rm DBC}= {\rm fixed} 
\hspace{3mm}
\Rightarrow
\hspace{3mm}
\left. \de q_{0,1} \right|_{\rm DBC} = 0 \, ,
\eeq
where the $|_{\rm DBC}$ refers to imposing Dirichlet boundary condition. Our surface 
contribution in eq. (\ref{eq:Sbd}) doesn't vanish under the imposition of 
this boundary condition. In the case when $\al=0$ (only Einstein-Hilbert gravity), 
then in order to have a sensible Dirichlet boundary value problem one has to add 
an extra boundary action. This is the well known Gibbon-Hawking-York term 
\cite{York:1986lje,Gibbons:1978ac,Brown:1992bq}, 
which in mini-superspace reduces to
\beq
\label{eq:GHYact}
S_{\rm GHY} = - \frac{V_3}{16 \pi G} \left. \frac{3 q \dot{q}}{N_c} \right|_0^1
= - \frac{V_3}{16 \pi G} 
\left(
\frac{3 q_1 \dot{q}_1}{N_c} - \frac{3 q_0 \dot{q}_0}{N_c}
\right) \, .
\eeq
On varying the $S_{\rm GHY}$ action it is noticed 
that it cancels the $\de \dot{q}$ terms at the boundary in eq. (\ref{eq:Sbd})
for $\al=0$. It therefore sets up a successful imposition dirichlet boundary 
condition, atleast for the Einstein-Hilbert gravity part of theory, thereby leading to 
a consistent solution to equation of motion. 

But the same thing can not be implemented for the Gauss-Bonnet sector 
of gravitational surface terms. They will be proportional 
to $f(\dot{q})\de \dot{q}$, where $f(\dot{q}) = \left(1/2N_c + \dot{q}^2/8N_c^3\right)$.
In principle one can construct a possible surface term for the Gauss-Bonnet sector.
\beq
\label{eq:GBsec_surfConst}
\left. S_{GB} \right|_{\rm bdy} = \left. F(q, \dot{q}) \right|_0^1 \, .
\eeq
During the process variation of action with respect to $q$ to compute 
equation of motion, this surface term on variation will lead to 
\beq
\label{eq:Fvar}
\ep \left. \left(
\frac{\pt F}{\pt q} \de q + \frac{\pt F}{\pt \dot{q}} \de \dot{q} 
\right) \right|_0^1 \, .
\eeq
Then in order to cancel the surface contribution proportional to $\al$
in eq. (\ref{eq:Sbd}), we notice that implies 
\bea
\label{eq:Fwork1}
&&
\ep \left.\frac{\pt F}{\pt \dot{q}} \de \dot{q} \right|_0^1
+ \frac{\ep V_3}{16 \pi G} 24 \al 
\left.\left\{ 
\left(\frac{k}{2N_c} + \frac{\dot{q}^2}{8N_c^3} \right) \de \dot{q} \right\}  \right|_0^1 = 0 \, ,
\notag \\
\Rightarrow 
&&
F(q, \dot{q}) = - \frac{V_3}{16 \pi G} 24 \al 
\left(\frac{k \dot{q}}{2N_c} + \frac{\dot{q}^3}{24 N_c^3} \right) + g(q) \, .
\eea
As the Gauss-Bonnet surface part in eq. (\ref{eq:Sbd}) doesn't have 
any term proportional to $\de q_0$ or $\de q_1$, so this implies 
that $g'(q)=0$, which can be fixed to zero. Then the total boundary action is a summation of 
Gibbon-Hawking term from eq. (\ref{eq:GHYact}) and 
Gauss-Bonnet part coming from eq. (\ref{eq:Fwork1}). 
\beq
\label{eq:Sbd_dir}
S_{\rm surface} = S_{\rm GHY} - \frac{V_3}{16 \pi G} 24 \al 
\left(\frac{k \dot{q}}{2N_c} + \frac{\dot{q}^3}{24 N_c^3} \right) \, .
\eeq 
This when added to the boundary contributions coming from varying the 
bulk action results in complete cancelation of the terms proportional to $\al$. 
As a result it doesn't lead to any non-trivial contributions coming from 
Gauss-Bonnet sector. However, for Dirichlet boundary conditions the equation 
of motion can still be solved without any inconsistencies, but the gravitational path-integral 
will not have non-trivial features coming from Gauss-Bonnet sector of gravitational action.
In a sense if our motivation is look for situations where Gauss-Bonnet piece of 
gravitational action contribute non-trivially then DBC doesn't fall in the category.

\subsection{Mixed Boundary condition (MBC)}
\label{mixedBC}

After not being able to have a consistent boundary value problem with Neumann 
boundary conditions and lack of obtaining non-trivial effects in the case of Dirichlet 
boundary conditions, we next consider the situation with mixed 
boundary conditions where we specify $q$ at one end and $\dot{q}$ at another end. 
Similar mixed boundary conditions have also been investigated in 
\cite{Krishnan:2017bte,DiTucci:2019dji,DiTucci:2019bui,DiTucci:2020weq}, here 
inspired by their work we consider applying them in case of Gauss-Bonnet gravity. 

In this case there are two possibilities: 
\bea
\label{eq:BC_cases}
&& {\rm Case \,\, (a)}: {\rm Specify\,\,} q_0 {\rm \,\, and\,\,} \dot{q}_1 
\Rightarrow \de q_0 = \de \dot{q}_1 =0 \, ,
\notag \\
&& {\rm Case \,\, (b)}: {\rm Specify\,\,} \dot{q}_0 {\rm \,\, and\,\,} q_1 
\Rightarrow \de q_1 = \de \dot{q}_0=0 \, .
\eea
We will consider each of this cases individually in more detail later in paper. 
But first we study the boundary action that is needed for each of these. 
The surface action for each of these is 
\bea
\label{eq:Surfact_a}
&&
S_{\rm surface}^{(a)} = \frac{V_3}{16 \pi G}
\biggl[
\frac{3 q_0 \dot{q}_0}{N_c} 
+ 24 \al \left(\frac{k\dot{q}_0}{2N_c} + \frac{\dot{q}_0^3}{24 N_c^3} \right)
\biggr] \, , 
\\
\label{eq:Surfact_b}
&&
S_{\rm surface}^{(b)} = -\frac{V_3}{16 \pi G}
\biggl[
\frac{3 q_1 \dot{q}_1}{N_c} 
+ 24 \al \left(\frac{k\dot{q}_1}{2N_c} + \frac{\dot{q}_1^3}{24 N_c^3} \right)
\biggr] \, .
\eea
During the computation of equation of motion, each of them can be varied and added to the 
boundary action in eq. (\ref{eq:Sbd}). This will result in 
\bea
\label{eq:SFact_a_var}
&&
S_{\rm bdy} + \de S_{\rm surface}^{(a)}
= \frac{\ep V_3}{16 \pi G} \biggl[
\frac{3}{N_c} \left(q_1 \de \dot{q}_1 + \dot{q}_0 \de q_0 \right)
+ 24 \al \left(\frac{k \de \dot{q}_1}{2N_c} + \frac{\dot{q}_1^2\de \dot{q}_1}{8N_c^3} \right) 
\biggr] \, , \\
\label{eq:SFact_b_var}
&&
S_{\rm bdy} + \de S_{\rm surface}^{(b)}
= -\frac{\ep V_3}{16 \pi G} \biggl[
\frac{3}{N_c} \left(\dot{q}_1 \de q_1 + \dot{q}_0 \de q_0 \right)
+ 24 \al \left(\frac{k \de \dot{q}_0}{2N_c} + \frac{\dot{q}_0^2\de \dot{q}_0}{8N_c^3} \right)
\biggr] \, .
\eea
From this one immediately notices that in former case $(a)$ if we fix $q_0$ and $\dot{q}_1$
then RHS of eq. (\ref{eq:SFact_a_var}) vanishes. Similarly in the later case $(b)$
if we fix $q_1$ and $\dot{q}_0$ then the RHS of the eq. (\ref{eq:SFact_b_var}) 
vanishes. In this way the boundary value problem is well-posed. Moreover the total action 
of theory is 
\bea
\label{eq:ToTact_a_mix}
S_{\rm tot}^{(a)} = S + S_{\rm surface}^{(a)} &=& 
\frac{V_3}{16 \pi G} \int {\rm d}t \biggl[
(6 k - 2\Lam q) N_c + \frac{3 \dot{q}^2}{2N_c}
+ \frac{3q \ddot{q}}{N_c} 
\biggr] 
\notag \\
&&
+ \frac{V_3}{16 \pi G}
\biggl[
\frac{3 q_0 \dot{q}_0}{N_c} 
+ 24 \al \left(\frac{k \dot{q}_1}{2N_c} + \frac{\dot{q}_1^3}{24 N_c^3} \right) \biggr] \, ,
\\
\label{eq:ToTact_b_mix}
S_{\rm tot}^{(b)} = S + S_{\rm surface}^{(b)} &=& 
\frac{V_3}{16 \pi G} \int {\rm d}t \biggl[
(6 k - 2\Lam q) N_c + \frac{3 \dot{q}^2}{2N_c}
+ \frac{3q \ddot{q}}{N_c}
\biggr] 
\notag \\
&&
- \frac{V_3}{16 \pi G}
\biggl[
\frac{3 q_1 \dot{q}_1}{N_c} 
+ 24 \al \left(\frac{k \dot{q}_0}{2N_c} + \frac{\dot{q}_0^3}{24 N_c^3} \right) \biggr] \, .
\eea
In each of these cases one can compute the \textit{momentum} corresponding to 
field variable $q(t)$ by varying the bulk Lagrangian with respect to $\dot{q}$.
This is given by
\beq
\label{eq:Spot_mom}
\pi = \frac{\de {\cal L}}{\de \dot{q}} = \frac{3 \dot{q}}{N_c} \, .
\eeq
It should be noted the bulk momentum in both the cases is same. 

The variational problem in the two cases is well-posed resulting in equation 
of motion whose solution can be found consistently. 
The solution to equation of motion in each of these cases is given by
\bea
\label{eq:qsol_a}
&&
q^{(a)}(t) = \frac{\Lam N_c^2}{3} t^2 + \left(\dot{q}_1 - \frac{2 \Lam N_c^2}{3} \right) t + q_0 \, ,
\\
\label{eq:qsol_b}
&&
q^{(b)}(t) = \frac{\Lam N_c^2}{3} t^2 + \dot{q}_0 (t-1) + \left(q_1 - \frac{\Lam N_c^2}{3} \right) \, .
\eea
These solution can be plugged back in corresponding action of theory 
in eq. (\ref{eq:ToTact_a_mix} \& \ref{eq:ToTact_b_mix})
to obtain action for the lapse $N_c$. The lapse action for the two cases is given by,
\bea
\label{eq:SNact_a_mix}
&&
S_{\rm tot}^{(a)} = \frac{V_3}{16 \pi G} \biggl[
6k N_c + \frac{3 \dot{q}_1 (2 q_0 + \dot{q}_1)}{2 N_c} 
- (2 q_0 + \dot{q}_1) N_c \Lam + \frac{2 N_c^3 \Lam^2}{9} 
+ \frac{\al \dot{q}_1}{N_c} \left(12k + \frac{\dot{q}_1^2}{N_c^2} \right)
\biggr] \, , 
\\
\label{eq:SNact_b_mix}
&&
S_{\rm tot}^{(b)} = \frac{V_3}{16 \pi G} \biggl[
6k N_c + \frac{3 \dot{q}_0 (\dot{q}_0 - 2 q_1)}{2 N_c} 
+ (\dot{q}_0 - 2 q_1) N_c \Lam + \frac{2 N_c^3 \Lam^2}{9} 
- \frac{\al \dot{q}_0}{N_c} \left(12k + \frac{\dot{q}_0^2}{N_c^2} \right)
\biggr] \, .
\eea
The lapse action include non-trivial features coming from the Gauss-Bonnet 
sector of gravitational action, which arise in the case of MBC. 
In the following we will study these two cases in more detail.

\section{Transition Probability}
\label{TranProb}

Generically once the action of a theory is known at the classical level then 
it can be used in the path-integral to study the behaviour of the corresponding 
quantum theory. In the case investigated 
in present paper the well-known classical action of Einstein-Hilbert gravity is modified 
by inclusion of Gauss-Bonnet gravity terms which is topological in four spacetime dimensions.
Although such topological extensions doesn't affect dynamical evolution of fields 
at the classical level as has been noted in the previous section, but their presence 
play a crucial role in dictating the choice of boundary conditions. 

In the case of gravitational path-integral one can study a simpler situation 
by restricting oneself to the mini-superspace approximation. Within this approximation 
one can precisely ask the following question: what is the transition amplitude from one $3$-geometry to another?
Is it possible to address this directly in Lorentzian signature? 
and what is the role played by boundary conditions 
in the computation of this transition amplitude given by path-integral? 
The relevant quantity that we are interested in can be expressed in mini-superspace approximation
as follows (see \cite{Halliwell:1988ik,Feldbrugge:2017kzv} for the Euclidean 
gravitational path-integral in mini-superspace approximation)
\beq
\label{eq:Gamp}
G[{\rm bd}_0, {\rm bd}_1]
= \int_{0^+}^{\infty} {\rm d} N_c  \int_{{\rm bd}_0}^{{\rm bd}_1} {\cal D} q(t) \,\, 
\exp \left(\frac{i}{\hbar} S_{\rm tot} \right) \, , 
\eeq
where ${\rm bd}_0$ and  ${\rm bd}_1$ are initial and final boundary configurations respectively.
The path-integral over $q(t)$ is performed such that it respects those boundary conditions. 
For our present case the above path-integral will be analysed with 
mixed boundary conditions as discussed in 
eq. (\ref{eq:BC_cases}) in previous section. $S_{\rm tot}$ is the total action 
incorporating the appropriate boundary condition as given in 
(\ref{eq:ToTact_a_mix} and \ref{eq:ToTact_b_mix}) respectively. 
The original contour of integration for $N_c$ is $(0^+, \infty)$. 
This contour integral over $N_c$ will be computed  
using the technology of Picard-Lefschetz theory.   

We start by considering the fluctuations around the
solution to equation of motion, which has been obtained previously
respecting the boundary conditions.  
\beq
\label{eq:qdecomp}
q(t) = \bar{q}^{(a,b)}(t) +  \ep^\prime \sqrt{8\pi G} Q(t) \, ,
\eeq
where $\bar{q}^{(a,b)}(t)$ is the solution to equation of motion 
given in eq. (\ref{eq:qsol_a} \& \ref{eq:qsol_b}), 
$Q(t)$ is the fluctuation around the background $\bar{q}^{(a,b)}(t)$, 
and $\ep^\prime$ is the parameter to keep track of order of terms.
This decomposition can be plugged back in total action given in 
(\ref{eq:ToTact_a_mix} and \ref{eq:ToTact_b_mix}) and expanded to 
second-order in $\ep^\prime$. $Q(t)$ obeys similar set of boundary conditions 
as the background $\bar{q}^{(a,b)}(t)$:
\bea
\label{eq:QBC}
&& {\rm Case \,\, (a)}: {\rm Specify\,\,} Q_0 {\rm \,\, and\,\,} \dot{Q}_1 
\Rightarrow Q_0 = \dot{Q}_1 =0 \, ,
\notag \\
&& {\rm Case \,\, (b)}: {\rm Specify\,\,} \dot{Q}_0 {\rm \,\, and\,\,} Q_1 
\Rightarrow Q_1 = \dot{Q}_0=0 \, .
\eea
After imposing these boundary conditions on $Q$ and performing the expansion in powers of $\ep^\prime$ 
we notice that first order terms in $\ep^\prime$ vanish as $\bar{q}^{(a,b)}(t)$ satisfies equation of 
motion. The second order terms are non-vanishing. The series in $\ep^\prime$ stops at 
second order. The full expansion can be written as
\beq
\label{eq:Sexp_Q}
S^{(a,b)} = S^{(a,b)}_{\rm tot} - \frac{3 \ep^{^\prime2} V_3}{4N_c} \int_0^1 {\rm d}t \dot{Q}^2 \, ,
\eeq
where $S^{(a,b)}_{\rm tot}$ is given in eq. (\ref{eq:SNact_a_mix} \& \ref{eq:SNact_b_mix}). 
In the path-integral measure such a decomposition will imply 
\beq
\label{eq:Mes_dec}
\int {\cal D} q(t) \Rightarrow \int {\cal D} Q(t) \, .
\eeq
As the action in eq. (\ref{eq:Sexp_Q}) separates into a part independent of $Q$ and 
part quadratic in $Q$, therefore the path-integral over $Q$ can be performed 
independently of the rest. This path-integral over $Q$ is 
\beq
\label{eq:pathQ_sep}
F(N_c) = 
\underset{\text{Case (a)}}{
\boxed{
\int_{Q[0]=0}^{Q'[1]=0} 
{\cal D} Q(t)
}}
\hspace{3mm}
{\rm OR}
\hspace{3mm}
\underset{\text{Case (b)}}{
\boxed{
\int_{Q'[0]=0}^{Q[1]=0} 
{\cal D} Q(t) 
}}
\exp \left(
- \frac{3 i \ep^{^\prime2} V_3}{4 \hbar N_c} \int_0^1 {\rm d}t \dot{Q}^2
\right) \, .
\eeq
This path-integral is very similar to the path-integral for a free 
where the trajectories at end points are kept fixed. However, this one is 
slightly different as at one of the boundary we are fixing $\dot{Q}$. 
A similar path-integral over mixed boundary conditions was encountered in 
\cite{DiTucci:2020weq} where the authors have computed it in 
appendix of the paper. Following the footsteps in \cite{DiTucci:2020weq}
we note
\beq
\label{eq:FNc_mbc}
F(N_c) = \frac{1}{\sqrt{\pi i}} \, .
\eeq
The important point to note is that in case of 
mixed boundary conditions the above path-integrals leads a 
$N_c$-independent numerical factor, unlike in case of 
Dirichlet boundary conditions where the 
above path-integral is proportional to $N_c^{-1/2}$. 

Then our transition amplitude $G[{\rm bd}_0, {\rm bd}_1]$ becomes 
\beq
\label{eq:Gab_afterQ}
G[{\rm bd}_0, {\rm bd}_1]
= \frac{1}{\sqrt{\pi i}} \int_{0^+}^\infty {\rm d} N_c \,\, 
\exp \left(\frac{i}{\hbar} S^{(a,b)}_{\rm tot} \right) \, ,
\eeq
where $S^{(a,b)}_{\rm tot}$ is given in eq. (\ref{eq:SNact_a_mix} \& \ref{eq:SNact_b_mix}). 
Now the task is reduced to performing the contour integration over lapse $N_c$. 
Here we will make use of complex analysis and Picard-Lefschetz formalism 
to analyse this integral. We start by studying the various saddle points of the 
action $S^{(a,b)}_{\rm tot}$ appearing in the exponent.

\subsection{Saddle points}
\label{sadpot}

The saddle points of the action can be found using
\beq
\label{eq:sadpot_eq}
\frac{\pt S^{(a,b)}_{\rm tot}}{\pt N_c} = 0 \, .
\eeq
The important thing to note here is about the structure of $S^{(a,b)}_{\rm tot}$
in terms of $N_c$ which can be noticed from eq. (\ref{eq:SNact_a_mix} \& \ref{eq:SNact_b_mix}).
It has term proportional to $N_c^3$, $N_c$, $1/N_c$ and $1/N_c^3$. Setting 
$(8 \pi G) =1$ the structural form for $S^{(a,b)}_{\rm tot}$ can be written as
\beq
\label{eq:Struct_Sab}
S^{(a,b)}_{\rm tot} = \frac{V_3}{2} \biggl[
A^{(a,b)} N_c + \frac{B^{(a,b)}}{N_c} + \frac{2 N_c^3 \Lam^2}{9}
+ \frac{\al C^{(a,b)}}{N_c^3} 
\biggr] \, ,
\eeq
where
\bea
\label{eq:Aab}
A^{(a)} = 6k - (2 q_0 + \dot{q}_1) \Lam \, ,
\hspace{3mm}
&&
A^{(b)} = 6k + (\dot{q}_0 - 2 q_1) \Lam \, , 
\\
B^{(a)} = \frac{3 \dot{q}_1 (2 q_0 + \dot{q}_1)}{2} + 12 \al k \dot{q}_1\, ,
\hspace{3mm} 
&&
B^{(b)} = \frac{3 \dot{q}_1 (\dot{q}_0 - 2 q_1)}{2} - 12 \al k \dot{q}_0
\\
C^{(a)} = \dot{q}_1^3 \, , 
\hspace{3mm}
&&
C^{(b)} = - \dot{q}_0^3 \, .
\eea
This structure is largely same as in the case of Einstein-Hilbert gravity, except the 
emergence of new additional term proportional to $1/N_c^3$ which is coming from the 
Gauss-Bonnet sector. The presence of this new term give rise 
to additional saddle points which are absent in the case of Einstein-Hilbert gravity. 
\beq
\label{eq:SabDN}
\frac{\pt S^{(a,b)}_{\rm tot}}{\pt N_c} = 0 
\hspace{5mm}
\Rightarrow
\hspace{5mm}
A^{(a,b)} - \frac{B^{(a,b)}}{N_c^2} + \frac{2 N_c^2 \Lam^2}{3}
- \frac{3 \al C^{(a,b)}}{N_c^4} = 0\, .
\eeq
It can be seen from this that the saddle point equation is cubic in $N_c^2$, 
resulting in three pairs of roots. This cubic equation can be solved by the 
known methods of dealing with cubic polynomial equation. 
In particular if the cubic equation has real coefficients then the nature of roots 
can be determined by analysing the behaviour of the 
discriminant of cubic equation. Such a strategy is no longer 
valid if the coefficients are complex. 
The discriminant of cubic polynomial with real coefficients is given by
\bea
\label{eq:discCB}
\D = &&\left(A^{(a,b)}\right)^2\left(B^{(a,b)}\right)^2 + \frac{8 \Lam^2}{3} \left(B^{(a,b)}\right)^2
+ 12 \al \left(A^{(a,b)}\right)^3 \left(C^{(a,b)}\right)
\notag \\
&&
+ 36 \al \Lam^2 \left(A^{(a,b)}\right) \left(B^{(a,b)}\right) \left(C^{(a,b)}\right)
- 108 \al^2 \Lam^4 \left(C^{(a,b)}\right)^2 \, .
\eea
If $\D>0$ the cubic equation has three distinct real roots for $N_c^2$. If $D<0$ the equation has 
two complex-conjugate roots, and one real root for $N_c^2$. 
By defining variables 
\bea
\label{eq:varUV}
U &=& \frac{3}{4 \Lam^4} \left(A^{(a,b)} \right)^2 + \frac{3}{2\Lam^2} B^{(a,b)} \, ,
\notag \\
V &=& \frac{3}{4 \Lam^4} B^{(a,b)} A^{(a,b)} 
+ \frac{1}{4 \Lam^6} \left(A^{(a,b)}\right)^3 - \frac{9\al}{2 \Lam^2} C^{(a,b)} 
\eea
one can write the roots as
\bea
\label{eq:N_roots}
N_0^{\pm} = \pm \left(
y_{+} + y_{-} - \frac{1}{2\lam^2} A^{(a,b)}
\right)^{1/2} \, , 
\notag \\
N_1^\pm = \pm \left(
y_{+} \om + y_{-} \om^2 - \frac{1}{2\lam^2} A^{(a,b)}
\right)^{1/2} \, , 
\notag \\
N_2^\pm = \pm \left(
y_{+} \om^2 + y_{-} \om - \frac{1}{2\lam^2} A^{(a,b)}
\right)^{1/2} \, , 
\eea
where 
\beq
\label{eq:ypm}
y_\pm = \left(\frac{V}{2} \pm \sqrt{\frac{V^2}{4} - \frac{U^3}{27} } \right)^{1/3} \, .
\eeq
where $1$, $\om$ and $\om^2$ are the three roots of unity. These are the six saddle points that 
arise in this system. 

The boundary conditions decide the nature of $A^{(a,b)}$, $B^{(a,b)}$
and $C^{(a,b)}$. If they are real then one can compute the discriminant of the cubic 
equation whose behaviour dictates the kind of roots for $N_c^2$. 
We can collectively write the saddle point as 
$N_\sg^\pm$, where $\sg=0$, $1$, and $2$.
Corresponding to each of these saddle points we have a metric 
\beq
\label{eq:sadpot_met}
\left({\rm d}s^{(a,b)}_\sg\right)^2 = - \frac{N_\sg^2}{q^{(a,b)}(t)} {\rm d} t^2 
+ q^{(a,b)}(t) \left[
\frac{{\rm d}r^2}{1-kr^2} + r^2 {\rm d} \OM_2^2
\right] \, ,
\eeq
where $q^{(a,b)}(t)$ is given by eq. (\ref{eq:qsol_a} \& \ref{eq:qsol_b}). 
Note that it is $N_\sg^2$ that enters the metric, which implies that the 
metric is same for each pair $N_\sg^\pm$ of saddle points.
As long as $N_\sg^2$ is real and positive, we are in Lorentzian signature. 
When it is real and negative then it is Euclidean signature, as in those 
cases $N_\sg$ is imaginary. In cases when $N_\sg^2$ is complex, the 
spacetime has a mixed signature. Geometries become singular 
when $q^{(a,b)}(t) \to 0$. In this case the spacetime volume goes to 
zero. 

For each of these saddle points one has a corresponding 
on-shell action. As the saddle points will generically be complex in nature therefore their 
corresponding on-shell action will have a real and an imaginary part.
The momentum at the saddles can be computed using eq. (\ref{eq:Spot_mom}).
\beq
\label{eq:sadpot_Momab}
\pi^{(a,b)} = \frac{3 \dot{q}^{(a,b)}}{N_\sg^{\pm}} \, .
\eeq
By making use of solution to equation of motion given in (\ref{eq:qsol_a} \& \ref{eq:qsol_b})
one can compute the momentum at the end points. 
\bea
\label{eq:SDpot_piab}
&&
\pi_0^{(a)} = \frac{3}{N_\sg^{(a)\pm}} \left[\dot{q}_1 - \frac{2 \Lam N_\sg^{(a)2}}{3} \right] \, ,
\hspace{5mm}
\pi_1^{(a)} = \frac{3 \dot{q}_1}{N_\sg^{(a)\pm}} \, , 
\notag \\
&&
\pi_0^{(b)} = \frac{3 \dot{q}_0}{N_\sg^{(b)\pm}} \, ,
\hspace{5mm}
\pi_1^{(b)} = \frac{3}{N_\sg^{(b)\pm}} \left(\frac{2 \Lam N_\sg^{(b)2}}{3} + \dot{q}_0 \right) \, .
\eea
The crucial point to note here is that momentum at the boundaries can be 
complex if the saddle point $N_\sg$ is complex. This is interesting as it carries 
characteristics of \textit{tunneling} phenomena.

\section{$N_c$-integration via Picard-Lefschetz}
\label{Ninnt}

We then go forth to compute the $N_c$-integration. We will make use of 
Picard-Lefschetz (PL) theory to analyse the behavior of the integrand 
in the complex plane \cite{Witten:2010cx,Witten:2010zr,Basar:2013eka,Tanizaki:2014xba}. 
Along with PL theory we make use of WKB methods to compute the 
integral. For this we need the set of saddle points and collection of 
steepest descent/ascent paths associated with each saddle point. 
A saddle point is termed `\textit{relevant}' if the steepest ascent path 
emanating from it intersects the original contour integration. 
The original integration contour can then be distorted 
to lie along the steepest descent paths passing through 
\textit{relevant} saddle points. Instead of using the prescription of 
Wick-rotation to deform the contour, we follow the methods of 
PL-theory to choose a contour of integration uniquely, 
along which the integrand is absolutely convergent. 

The problem of performing path-integration is reduced to a task 
of computing thimbles (steepest descent paths) on a complex plane. 
In the following we will give a review of Picard-Lefschetz formalism.
We start by considering the path-integral in the following manner 
\beq
\label{eq:pathmock}
I = \int {\cal D}z(t) \, e^{i {\cal S}(z)/\hbar} \, ,
\eeq
where the exponent is functional of $z(t)$. In general the
integrand can be quite oscillatory and hence not an easy task to 
compute the integral. In flat spacetime the global symmetries of 
spacetime allow one to cast Lorentz group in to 
a compact rotation group under a transformation of time co-ordinate. 
This privilege doesn't exist in non-flat spacetimes. 
Such a transformation of time co-ordinate in flat spacetime 
leads to exponential damping of above integrand.
In PL theory one analytically continues both 
$z(t)$ and ${\cal S}(z)$ in to complex plane, and 
interprets ${\cal S}$ as an holomorphic functional of
$z(t)$. This implies that ${\cal S}$ satisfies a 
functional form of Cauchy-Riemann conditions
\begin{align}
\label{eq:CRfunc}
\frac{\de {\cal S}}{\de \bar{z}} = 0 
\Rightarrow
\begin{cases}
\frac{\de {\rm Re} {\cal S}}{\de x}
&= \frac{ \de {\rm Im} {\cal S}}{\de y} \, , \\
\frac{\de {\rm Re} {\cal S}}{\de y}
&= - \frac{ \de {\rm Im} {\cal S}}{\de x} \, .
\end{cases}
\end{align}
%

\subsection{Flow equations}
\label{floweq}

On writing the complex exponential as
${\cal I} = i {\cal S}/\hbar = h + iH$ and writing 
$z(t) = x_1(t) + i x_2(t)$, the downward flow 
is defined as
\beq
\label{eq:downFlowDef}
\frac{{\rm d} x_i}{{\rm d} \lam}
= - g_{ij} \frac{\pt h}{\pt x_j} \, ,
\eeq
where $g_{ij}$ is a metric defined on the complex manifold, 
$\lam$ is flow parameter and $(-)$ sign refers to downward flow. 
The steepest descent flow lines follow a trajectory 
dictated by above equation. They are also knowns as \textit{thimbles} 
(can be denoted by ${\cal J}_\sg$).
Steepest ascent flow lines are defined with a plus sign in front of 
$g_{ij}$ in the eq. (\ref{eq:downFlowDef}), and are denoted as ${\cal K}_\sg$.
Here $\sg$ refers to the saddle point to which these flow-lines are attached. 
The definition of flow lines immediately implies that the real part $h$ (also called 
Morse function) decreases monotonically as one moves 
away from the critical point along the steepest descent curves.
This can be seen by computing 
\beq
\label{eq:flowMonoDec_h}
\frac{{\rm d} h}{{\rm d} \lam}
= g_{ij} \frac{{\rm d} x^i}{{\rm d} \lam} \frac{\pt h}{\pt x_j} 
= - \left(\frac{{\rm d} x_i}{{\rm d}\lam}\frac{{\rm d} x^i}{{\rm d}\lam}\right)
\leq 0 \, .
\eeq
This generically holds for any Riemannian metric. However, in this 
paper for simplicity we assume $g_{z,z}=g_{\bar{z},\bar{z}}=0$
and $g_{z,\bar{z}}= g_{\bar{z},z}=1/2$. This leads to a simplified version of
flow equations 
\beq
\label{eq:simpflow}
\frac{{\rm d}z}{{\rm d} \lam} = \pm \frac{\pt \bar{\cal I}}{\pt \bar{z}} \, ,
\hspace{5mm}
\frac{{\rm d}\bar{z}}{{\rm d} \lam} = \pm \frac{\pt {\cal I}}{\pt z} \, .
\eeq
Using them it is easy to notice that the imaginary 
part of ${\rm Im} {\cal I}=H$ is constant along the flow lines. 
\beq
\label{eq:consHflow}
\frac{{\rm d}H}{{\rm d} \lam} 
= \frac{1}{2i} \frac{{\rm d} ({\cal I} - {\cal \bar{I}})}{{\rm d} \lam} 
= \frac{1}{2i} \left(
\frac{\pt {\cal I}}{\pt z} \frac{{\rm d} z}{{\rm d} \lam} 
- \frac{\pt \bar{\cal I}}{\pt \bar{z}}\frac{{\rm d}\bar{z}}{{\rm d} \lam}
\right) = 0 \, .
\eeq
This is a wonderful feature of flow-lines and can be used to determine the 
structure of flow-lines in the complex $N_c$-plane.
It is seen that the oscillatory integral becomes convergent and 
well-behaved along any of the steepest descent lines (thimbles). 
This motivates one check if it is possible to analytically deform 
the original integration contour to integration 
along either one thimble or a sum of thimbles. 
This is a true generalization of Wick-rotation. 

In the complex $N_c$-plane the flow equations corresponding to 
steepest descent (ascent) becomes the following in cartesian co-ordinates 
\begin{subequations}
\begin{align}
\label{eq:STdes}
& {\rm Descent} \Rightarrow 
& \frac{{\rm d} x_1}{{\rm d} \lam} = - \frac{\pt {\rm Re}{\cal I}}{\pt x_1} \, ,
\hspace{5mm}
&
\frac{{\rm d} x_2}{{\rm d} \lam} = - \frac{\pt {\rm Re}{\cal I}}{\pt x_2} \, , 
\\
\label{eq:STaes}
& {\rm Ascent} \Rightarrow 
& \frac{{\rm d} x_1}{{\rm d} \lam} = \frac{\pt {\rm Re}{\cal I}}{\pt x_1} \, ,
\hspace{5mm}
&
\frac{{\rm d} x_2}{{\rm d} \lam} = \frac{\pt {\rm Re}{\cal I}}{\pt x_2} \, .
\end{align}
\end{subequations}
It is noticed that the ${\rm Im} {\cal I}$ doesn't enter the 
flow equations as ${\rm Im} {\cal I}= {\rm const.}$ along the flow lines. 
Each saddle point has two steepest descent lines and two steepest 
ascent lines attached to it. The boundary conditions and the 
parameter values dictate the location of the saddles on the 
complex $N_c$-plane. Solving these flow equations can be 
sometimes hard as ${\cal I}$ can be complicated. However, it 
is possible to deal with them numerically. One can bypass solving them
entirely by making use of knowledge that $H$ is constant along them. 
This determines all the flow-lines. But to find out about the nature 
of flow lines one has to compute the gradient of first derivative 
(second order derivative of action at the saddle points).

\subsection{Choice of contour}
\label{choice}

Once the set of saddle points along with the set of steepest ascent/descent flow-lines 
associated with each saddle point are known, one can begin to find the 
new contour of integration to which the original integration contour will be deformed. 
The integral in complex $N_c$-plane is absolutely convergent along this 
new contour (for more detail see \cite{Witten:2010zr,Basar:2013eka,Feldbrugge:2017kzv}). 

In the complex $N_c$ plane the behavior of $h$ and $H$
determines the `allowed' regions (region where integral is well-behaved) and 
`forbidden' region (region where integral diverges). We label the former 
by $J_\sg$ while later is denoted by $K_\sg$, and as mentioned 
previously $\sg$ refers the saddle point. 
These regions have
$h(J_\sg) < h(N_\sg)$, while $h(K_\sg)>h(N_\sg)$. 
$h$ goes to $-\infty$ along the steepest descent lines and ends in a singularity, 
while along the steepest ascent contours $h\to+\infty$.
These lines usually intersect at only one point where they are both
well-defined. With a suitable choice of orientation one can write
\beq
\label{eq:JKorient}
{\rm Int} \left({\cal J}_\sg, {\cal K}_{\sg^\prime} \right) = \de_{\sg \sg^\prime} \, .
\eeq
The purpose is to write the integral over the original contour 
as an integral along the new contour which is sum of integrations 
done along Lefschetz thimbles. Schematically this can be expressed as
\beq
\label{eq:sadsum}
\mathbb{D} =(0^+,\infty) \Rightarrow {\cal C} = \sum_\sg n_\sg {\cal J}_\sg \, ,
\eeq
in a homological sense for some integers $n_\sg$ which will take 
value $0$ or $\pm1$ when accounting for orientation 
of contour over each thimble. This will also imply that 
$n_\sg = {\rm Int} ({\cal C}, {\cal K}_\sg) = 
{\rm Int} (\mathbb{D}, {\cal K}_\sg)$. As the intersection number is 
topological and doesn't change if we deform the contour, therefore 
the necessary and sufficient condition for a thimble ${\cal J}_\sg$ to be relevant 
is that the steepest ascent curve from the corresponding 
saddle point intersects the original integration domain $\mathbb{D}$. 
The integration contour is chosen to lie in the region $J_\sg$
(which is the `allowed' region) and follow the contour trajectory dictated 
by the steepest descent paths \cite{Feldbrugge:2017kzv}.
In this circumstance there is no hindrance in smoothly sliding the 
intersection point along the ${\cal K}_\sg$ to the \textit{relevant}
saddle point. 

Once the original integration contour is deformed to a sum over integration 
done along various {\it relevant} thimbles then we have 
\beq
\label{eq:sumOthim}
I = \int_{\cal C} {\rm d} z(t) e^{i S[z]/\hbar}
= \sum_\sg n_\sg \int_{{\cal J}_\sg} {\rm d}z(t)
e^{i S[z]/\hbar} \, .
\eeq
It is common that in such process more than one thimble 
contributes to integration, resulting in interference of contributions 
coming from various thimbles. This is feature of performing 
complex integration via Picard-Lefschetz methodology. 
The integration along each of the thimbles is absolutely convergent if
\beq
\label{eq:absconvg}
\biggl| \int_{{\cal J}_\sg} {\rm d} z(t)
e^{i S[z]/\hbar} \biggr| \leq
\int_{{\cal J}_\sg} \lvert {\rm d} z(t) \rvert
\lvert e^{i S[z]/\hbar} \rvert
= \int_{{\cal J}_\sg} \lvert {\rm d} z(t) \rvert e^h(z) < \infty \, .
\eeq
If we denote the length along the contour path as $l= \int \lvert {\rm d} z(t) \rvert$,
then convergence of above integral require that $e^h \sim 1/l$ as $l\to \infty$.
The original integration hence can be analytically deformed into a sum of absolutely
convergent integrals along various Lefschtez thimbles passing through {\it relevant} 
saddle points. If one does an expansion 
in $\hbar$ then to leading order we get the following
\beq
\label{eq:LDordI}
I = \int_{\cal C} {\rm d} z(t) e^{i S[z]/\hbar}
= \sum_\sg n_\sg e^{i H(N_\sg)} \int_{{\cal J}_\sg} {\rm d}z(t) e^{h} 
\approx \sum_\sg n_\sg e^{i S[N_\sg]/\hbar} \left[{\cal A}_\sg 
+ {\cal O}(\hbar) \right] \, ,
\eeq
where ${\cal A}_\sg$ is the contribution coming after performing a 
gaussian integration around the saddle point $N_\sg$. 

\subsection{Flow directions}
\label{flowDir}

The direction of flow lines either emanating from the saddles or going into it can be 
determined analytically (to some extent) by expanding the $N_c$-action of theory 
given in eq. (\ref{eq:SNact_a_mix} \& \ref{eq:SNact_b_mix}) 
around the saddle points given in eq. (\ref{eq:N_roots}). 
If we write $N_c = N_\sg + \de N$ (where $N_\sg$ is any saddle point of action), 
then the action has a power series expansion in $\de N$.
\beq
\label{eq:NexpSad}
S^{(0)} = S^{(0)}_\sg + \left. \frac{{\rm d}S^{(0)}}{{\rm d}N_c} \right|_{N=N_\sg} \de N 
+ \frac{1}{2} \left. \frac{{\rm d}^2 S^{(0)}}{{\rm d}N_c^2} \right|_{N=N_\sg} \left(\de N\right)^2 
+ \cdots \, .
\eeq
The first order terms will vanish identically by definition.  

The second order terms can be computed directly from the action 
in eq. (\ref{eq:SNact_a_mix} \& \ref{eq:SNact_b_mix}), by just taking double-derivative with respect to $N_c$.
From this the direction of flow-lines can be determined. 
One should recall that the imaginary part of exponential $iS$ (or $H$) is constant along the flow lines.
This implies that ${\rm Im} \left[iS - i S (N_s) \right]=0$. 
The second variation at the saddle point can be written as 
${\rm d}^2 S^{(0)}/{\rm d}N_c^2 = re^{i \rho}$, 
where $r$ and $\rho$ depends on boundary conditions. 
Near the saddle point the change in $H$ will go like 
\beq
\label{eq:changeH}
\D(H) \propto i 
\left(\left. \frac{{\rm d}^2 S^{(0)}}{{\rm d}N^2} \right|_{N_\sg}\right) \left(\de N_c\right)^2
\sim n_\sg^2 e^{i\left(\pi/2 + 2\ta_\sg + \rho_\sg \right)} \, ,
\eeq
where we write $\de N = n_\sg e^{i \ta_\sg}$ and $\ta_\sg$ is the direction of flow lines
at the corresponding saddle point. 
Given that the imaginary part $H$ remains constant along the 
flow lines, so this means 
\beq
\label{eq:flowang}
\ta_\sg = \frac{(2k-1)\pi}{4} - \frac{\rho_\sg}{2} \, ,
\eeq
where $k \in \mathbb{Z}$. 

For the steepest descent and ascent flow lines, their 
corresponding $\ta_\sg^{\rm des/aes}$ is such that the phase for 
$\D H$ correspond to $e^{i\left(\pi/2 + 2\ta_\sg + \rho_\sg \right)} = \mp1$. This implies
\beq
\label{eq:TaDesAes}
\ta_\sg^{\rm des} = k \pi + \frac{\pi}{4} - \frac{\rho_\sg}{2} \, ,
\hspace{5mm}
\ta_\sg^{\rm aes} = k \pi - \frac{\pi}{4} - \frac{\rho_\sg}{2} \, .
\eeq
These angles can be computed numerically for the given boundary conditions
and for gravitational actions.

\subsection{Saddle-point approximation}
\label{sadPtApp}

Once the set of saddle points, flow directions and steepest descent/ascent paths 
associated with them (denoted by ${\cal J}_\sg/{\cal K}_\sg$ respectively)
are known, it is then easy to find the \textit{relevant} saddle points. 
A saddle point is termed \textit{relevant} if the steepest ascent path emanating from it 
intersect with the original contour of integration. In the current case the original 
integration contour is $(0^+, \infty)$. 
The original integration contour then becomes sum over the contribution 
coming from all the Lefschetz thimbles passing through {\it relevant} saddle points.
We can then do saddle-point-approximation to compute the 
transition amplitude in eq. (\ref{eq:Gab_afterQ}). In the 
$\hbar\to0$ limit we have
\beq
\label{eq:Gabapprox}
G[{\rm bd}_0,{\rm bd}_1] 
\approx
\frac{1}{\sqrt{\pi i}} \sum_\sg n_\sg 
\exp\left[\frac{i}{\hbar} S_{\rm tot}^{(a,b)} \left(N_\sg \right) \right]
\int_{{\cal J}_\sg}
{\rm d}N_c \exp \left[
\frac{i}{\hbar} \left(S_{\rm tot}^{(a,b)}\right)_{N_cN_c} (N_c - N_\sg)^2
\right] \, ,
\eeq
where we consider only the leading order term in $\hbar$. Here 
$N_\sg$ are the {\it relevant} saddle points for the various boundary conditions 
given in eq. (\ref{eq:N_roots}), $S_{\rm tot}^{(a,b)} \left(N_\sg \right)$ is the 
on-shell action which can be computed from eq. (\ref{eq:SNact_a_mix}
\& \ref{eq:SNact_b_mix}) for the {\it relevant} saddles. $\left(S_{\rm tot}^{(a,b)}\right)_{N_cN_c}$ is the 
second variation of the action with respect to $N_c$ computed at the 
{\it relevant} saddle points. 

On writing $N-N_\sg=n e^{i \ta_\sg}$, where $\ta_\sg$ is the angle the Lefschetz thimble make 
with the real $N$-axis while $\sg$ corresponds to {\it relevant} saddle point. 
Then the above integration can be performed easily. 
It gives the following 
\bea
\label{eq:GabTrans}
G[q_0, q_1] = \frac{1}{\sqrt{\pi i}}  
\sum_\sg n_\sg 
\lvert \left(S_{\rm tot}^{(a,b)}\right)_{N_cN_c} \rvert^{-1/2}
\exp\left[i \ta_\sg + \frac{i}{\hbar} S_{\rm tot}^{(a,b)}\left(N_\sg\right) \right]\, .
\eea
In the next section we will make use of it and apply it to the case of 
no-boundary proposal of Universe to compute the transition amplitude.

\section{No-boundary Universe}
\label{nbu}

This is special boundary condition where the Universe start from nothing. 
In the current situation this implies $q_0=0$, implying that Universe started with a 
zero scale factor $a$ 
\cite{Hawking:1981gb,Feldbrugge:2017kzv,Feldbrugge:2017fcc,Feldbrugge:2017mbc,DiTucci:2019dji,DiTucci:2019bui}. 
In case $(a)$ where we
specify field $q_0$ at one end point while its first derivative $\dot{q}_1$ at another end point
this immediately leads to a simplified $N_c$ action. However, at the final 
boundary, following the solution to equation of motion from eq. (\ref{eq:qsol_a})
we also have a relation
\beq
\label{eq:q1a_q1p}
\dot{q}_1^{(a)} = q_1^{(a)} + \frac{\Lam N_c^2}{3} \, .
\eeq
This allow us to express the derivative of field in terms of field value at the final 
boundary. This is useful as one can study the problem by doing the analysis in terms of $q_1$ 
where we require that for physical reasons $q_1>0$. A real and positive $q_1$ 
immediately implies a possibly complex $\dot{q}_1^{(a)}$ if $N_c^2$ is complex. 
On the other hand this also means that if $q_1$ has a fixed real positive value and the 
number of {\it relevant} saddle points are more that one, then it will imply that 
at the final boundary $\dot{q}_1^{(a)}$ will have multiple values. This is 
contradictory to our initial requirement that in case $(a)$ $\dot{q}_1^{(a)}$ is 
fixed at final boundary, and implies that the Universe at final time 
has multiple values of $\dot{q}_1^{(a)}$.

In the case $(b)$ on other hand the boundary conditions require fixing 
$\dot{q}_0^{(b)}$ and $q_1^{(b)}$.
This implies that at $t=0$ by following the solution \ref{eq:qsol_b}, if 
the Universe started from nothing ($q^{(b)}(t=0)=0$), then 
it leads to a relationship between $\dot{q}_0^{(b)}$ and $q_1^{(b)}$
\beq
\label{eq:q1inq0p}
\dot{q}_0^{(b)} = q_1^{(b)} - \frac{\Lam N_c^2}{3}   \, .
\eeq
Here if at final boundary $q_1^{(b)}>0$ (real and positive)
then the initial $\dot{q}_0^{(b)}$ could be complex when $N_c^2$ is complex. 
Moreover, when the number of {\it relevant} saddle points are more than 
one, then the final geometry is seen to arising from 
superposition multiple initial geometries. This is quite possible
and doesn't lead to a contradiction unlike in case $(a)$. 
In the following we will study this particular scenario in more detail. 
We can plug the relation in eq. (\ref{eq:q1inq0p}) in the action for $N_c$ for case $(b)$ 
in eq. (\ref{eq:SNact_b_mix}) to obtain the $N_c$-action for the no-boundary proposal. 
\bea
\label{eq:nbu_csb_act}
&&
S_{\rm tot}^{(b)} = \frac{V_3}{16 \pi G} \biggl[
6 k N_c + \frac{\{(\al \Lam -3) q_1 + 12 k \al\}N_c \Lam}{3} 
+ \frac{(9+2\al \Lam) q_1^2 + 24 k \al q_1}{2N_c} 
\notag \\
&&
+ \frac{(3+2\al\Lam) \Lam^2 N_c^3}{54}
+ \frac{\al q_1^3}{N_c^3} \biggr] \, .
\eea
It should be noted that if we set $\al=0$ then we get the action for the 
no-boundary Universe in case of pure Einstein-Hilbert gravity. 
We note that this residual action is bit different from the action 
that one obtains in the case of dirichlet boundary conditions 
\cite{Feldbrugge:2017kzv,Feldbrugge:2017fcc,Feldbrugge:2017mbc}.
This is because we used mixed boundary conditions to arrive at the 
action in eq. (\ref{eq:nbu_csb_act}). 
The saddle point equation correspondingly is
\bea
\label{eq:sadpt_Nc_b}
{\rm Case\,\,(b):} \hspace{3mm} 
&&
\frac{\Lam^2(3+2\al\Lam)}{18} N_c^6
+ \left\{6k - q_1 \Lam + \frac{\al \Lam (12k + q_1 \Lam)}{3} \right\} N_c^4
\notag \\
&&
- \left\{12 k \al q_1 + \left(\frac{9}{2} + \al \Lam \right) q_1^2 \right\} N_c^2
- 3 \al q_1^3 =0 \, .
\eea
The interesting thing to note here is that in case $(b)$ there exist a $q_1$ for which the 
coefficient of $N_c^4$ in eq. (\ref{eq:sadpt_Nc_b}) can vanishe. 
This will offer some simplification in the expressions for saddle points. 

The saddle-point equation is cubic in $N_c^2$ with real coefficients. Its nature 
of roots can be decided based by analysing the behaviour of its discriminant 
in the parameter space of couplings and boundary value $q_1$. 
It is seen that for positive $k$, $\Lam$, and $\al$ the  
discriminant is always positive for $q_1 \geq 0$. This is 
interesting as it quickly implies that the saddle point equation has three 
distinct real roots for $N_c^2$. Also as $\al$ and $q_1$ are positive, 
the zeroth-order term in $N_c^2$ in the saddle-point equation 
is positive. This means that the product of three roots has to be positive.
It leads to two possibilities: either all roots for $N_c^2$ are positive or
one is positive and other two are negative. However, as the coefficient of 
$N_c^2$ is negative so this immediately implies the later case
with one positive root and two negative roots for $N_c^2$. 
This means that we have two saddle points lying on real-axis in complex $N_c$-plane 
(one positive and one negative); while four saddle points lie
on imaginary axis in complex $N_c$-plane (two of them in positive 
imaginary axis, while other two in negative imaginary axis). 
It is worth stating here that saddle point where $N_c^2>0$ correspond to the usual 
Lorentzian geometry with a Lorentzian evolution, as can be seen from eq. (\ref{eq:sadpot_met}).
On the other hand the saddle point where $N_c^2<0$ correspond to Euclidean geometries. 

To workout the transition amplitude in eq. (\ref{eq:GabTrans}) one also needs to know 
the second variation of the above action. These will subsequently be needed to 
determine the direction of Lefschetz thimbles at the various {\it relevant} saddle points.
The second variation is given by
\beq
\label{eq:S2var_b}
\left(S^{(b)} \right)_{,N_c N_c} = \frac{V_3}{16\pi G} \left[
\frac{(3+2\al \Lam) \Lam^2 N_c}{9} 
+ \frac{(9+2\al \Lam) q_1^2 + 24 k \al q_1}{N_c^3}
+ \frac{12 \al q_1^3}{N_c^5}
\right] \, .
\eeq
At real saddle points the second variation is also real, while when 
$N_c$ is complex then the second variation will also be complex. 

The nature of \textit{relevance} of each of these saddle-points depends on the 
parameter values and whether the steepest ascent path emanating from them 
intersects the original integration contour. In principle this seems like a well-defined 
way of finding out the \textit{relevance} of saddle points. However, in practice 
often the action has large amount of symmetry. Due to this there is degeneracy 
between steepest ascent and steepest descent curves. It means that the
steepest ascent curve from one saddle point overlaps with the steepest descent 
curve from another saddle point. To lift these degeneracy one can add a small 
perturbation in the $N_c$ action which helps in breaking symmetry. Lifting this 
degeneracy also aid us to correctly locate the \textit{relevant} saddle points. 

We will consider a numerical example to investigate the state of art once 
parameters are fixed to some value. For numerical analysis and to lift 
degeneracy of the system we consider adding a small perturbation to the 
$N_c$ action 
\beq
\label{eq:pertact}
S_{\rm pert} = \frac{i V_3 \ep^{\prime\prime} N_c}{16\pi G} \, ,
\eeq
where $\ep^{\prime\prime}$ is a small parameter. Notice that the perturbation is imaginary in nature
and it is somewhat reminiscent to $+i\ep$-prescription in standard 
flat spacetime field theory. In a sense we are inspired by the Feynman's 
$+i\ep$-prescription to choose this perturbation.  

For purpose of better understanding the system, we pick up an example. 
We consider the value of parameters:
$k=1$, $\Lam=3$, $\al=2$ and $\ep^{\prime\prime}=10^{-2}$ (we have set 
$8\pi G=1$). For $k=1$ the volume $V_3 = \pi^2$ which can be computed 
using eq. (\ref{eq:volDm1}). We compute the saddle points following the 
eq. (\ref{eq:sadpt_Nc_b}). As discussed previously it is seen that the cubic
equation in $N_c^2$ has three distinct roots: $N_0^2$, $N_1^2$ and $N_2^2$. 
As expected the ${\rm Re} (N_0^2)>0$, ${\rm Re} (N_1^2)<0$ and ${\rm Re} (N_2^2)<0$
for all $q_1$.
Each of these roots for $N_c^2$ gets a small imaginary part due to the perturbation,
which thereby give rise to a small deviation in the saddle points value
\bea
\label{eq:sadDev}
&&
{\rm Im} \, \left(\de N_0^\pm\right) = \mp \nu_0 \, , \hspace{5mm} \nu_0>0 \, ,
\notag \\
&&
{\rm Re}  \, \left(\de N_1^\pm\right) = \pm \nu_1 \, ,\hspace{5mm} \nu_1>0 \, ,
\notag \\
&&
{\rm Re}  \, \left(\de N_2^\pm\right) = \pm \nu_2 \, ,\hspace{5mm} \nu_2>0 \, .
\eea

\begin{figure}[h]
\centerline{
\vspace{0pt}
\centering
\includegraphics[width=2.7in]{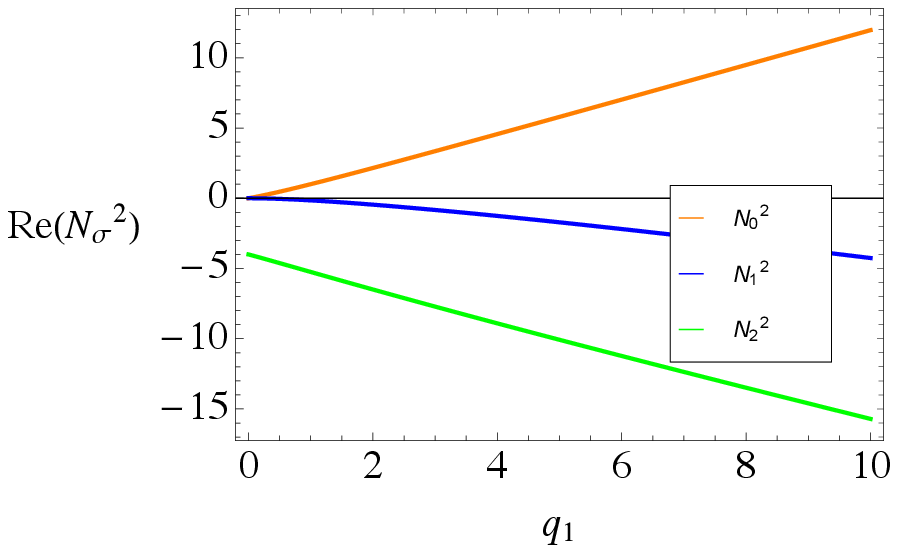}
\hspace{4mm}
\includegraphics[width=3in]{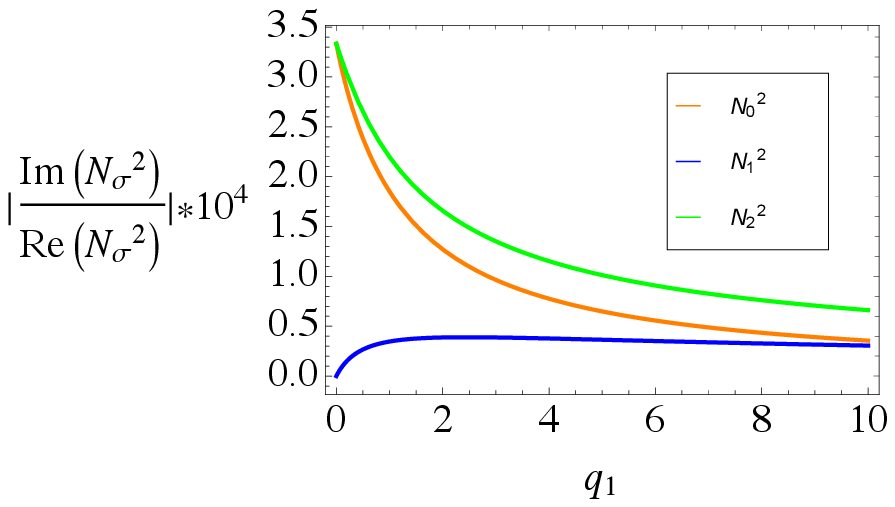}
}
\caption[]{
Here we analyse the real and imaginary part of the saddle points ($N_0^2$, $N_1^2$ and $N_2^2$)
for various values of $q_1$. In this numerical example we consider 
$k=1$, $\Lam=3$, $\al=2$ and $\ep^{\prime\prime}=10^{-2}$. There is a small imaginary 
part which comes due to the perturbation added to the total $N_c$-action.
As discussed the real part of $N_0^2$ remains positive, while 
real part of $N_1^2$ and $N_2^2$ remains always negative. 
In the plot on left we show the real part of various $N_\sg^2$, while 
on the right plot we see the behavior of $\lvert {\rm Im} (N_\sg^2)/ {\rm Re} (N_\sg^2) \rvert$
as a function of $q_1$. For the plot on right we have scaled the value by $10^4$
to plotting purpose.
}
\label{fig:nbu_sadpt}
\end{figure}

In figure \ref{fig:nbu_sadpt} we plot the real and imaginary part 
of these roots as a function of $q_1$ (the value of $q$ at $t=1$). 
%
%
To determine the nature of saddle points (\textit{relevant} or \textit{irrelevant})
one has to study the steepest descent/ascent flow lines corresponding 
to each saddle point. These flow lines can be drawn 
by exploiting the knowledge that along these lines $H(N_c) = H(N_\sg)$. 
In the absence of perturbation in (\ref{eq:pertact}) there will be 
some degeneracy in the sense that steepest descent line of one saddle 
will overlap with the steepest ascent line from another saddle. The addition 
of perturbation helps in lifting this degeneracy. In order to find the 
{\it relevance} of saddle points it is also crucial to analyse the nature of 
Morse-function $h$ at each of these saddle points. 
Picard-Lefschetz theory dictates that {\it relevant} saddle points 
must be reached by flowing down from the original integration 
contour via steepest ascent paths. This will immediately imply
that $h<0$ at the {\it relevant} saddle points. A complex action 
bypasses this rigid constraint though. However, in our present case 
this is not possible. 

\begin{figure}[h]
\centerline{
\vspace{0pt}
\centering
\includegraphics[width=3.1in]{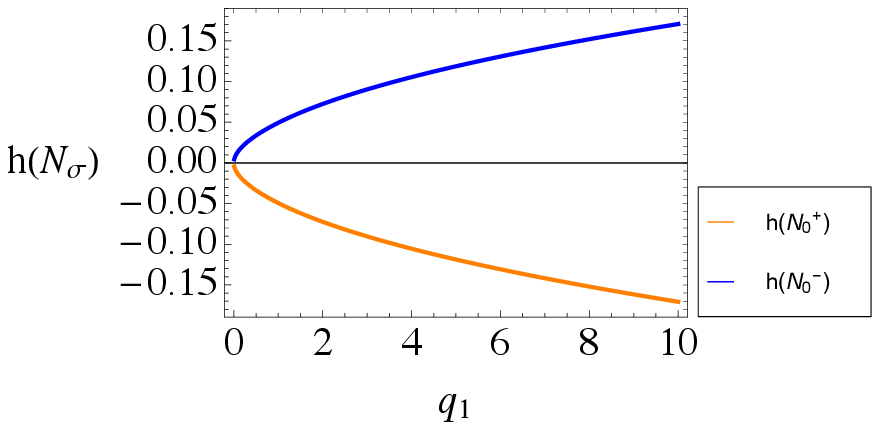}
\hspace{3mm}
\includegraphics[width=3in]{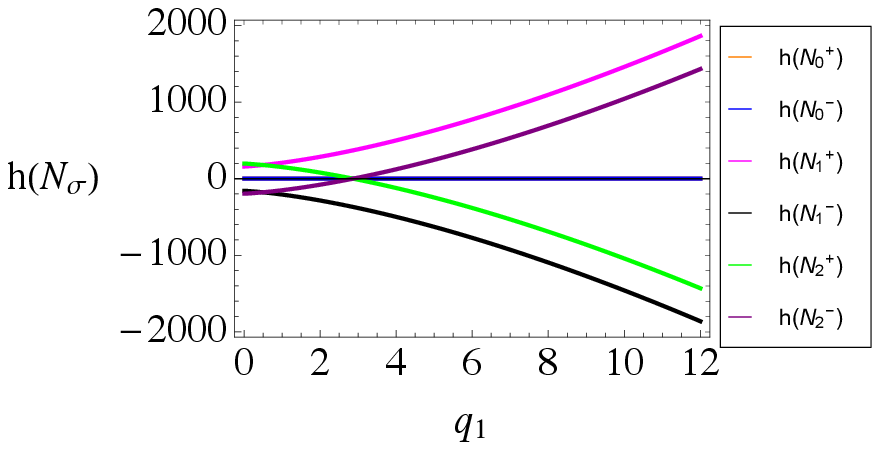}
}
\caption[]{
Plotting Morse-function $h$ for various saddle point against $q_1$. 
For this we consider parameter values $k=1$, $\Lam=3$, $\al=2$ and $\ep^{\prime\prime}=10^{-2}$. 
From the plots we notice that only $h(N_0^-)$ and $h(N_1^-)$ remains always negative. 
$h(N_2^\pm)$ change sign after a certain threshold $q_1^{\rm th}$. 
}
\label{fig:hsad}
\end{figure}

If we plot $h(N_\sg)$ against $q_1$ we notice that for some of saddle 
points $h$ changes sign as $q_1$ is varied. This is shown in 
figure. \ref{fig:hsad}. 
%
%
For the present situation there are six saddle points. Only those which 
can be reached by flowing downward along the steepest ascent lines 
from the original integration contour are {\it relevant}. 
The saddle points $N_0^+$ and $N_0^-$ lie in 
lower-right and upper-left quadrant respectively. Only the former 
can be reached via steepest ascent lines from original integration 
contour and hence is {\it relevant}. The saddles 
$N_1^+$ and $N_1^-$ lie in upper-right and lower-left
quadrant respectively. Both these saddle-point can be reached from 
original integration contour by flowing downward along the steepest 
ascent flow lines. However only the later lie in 
allowed region with corresponding $h<0$ and is favourable. 
$N_1^-$ is therefore {\it relevant}. 
The saddle points $N_2^+$ and $N_2^-$ lie in 
lower-right and upper-left quadrant respectively.
Both are {\it irrelevant}: the former can't be reached via a steepest ascent 
path from original integration contour while later has $h>0$. 
So out of six saddle-points only two of them are {\it relevant}:
$N_0^+$ and $N_1^-$. 
%
\begin{figure}[h]
\centerline{
\vspace{0pt}
\centering
\includegraphics[width=4in]{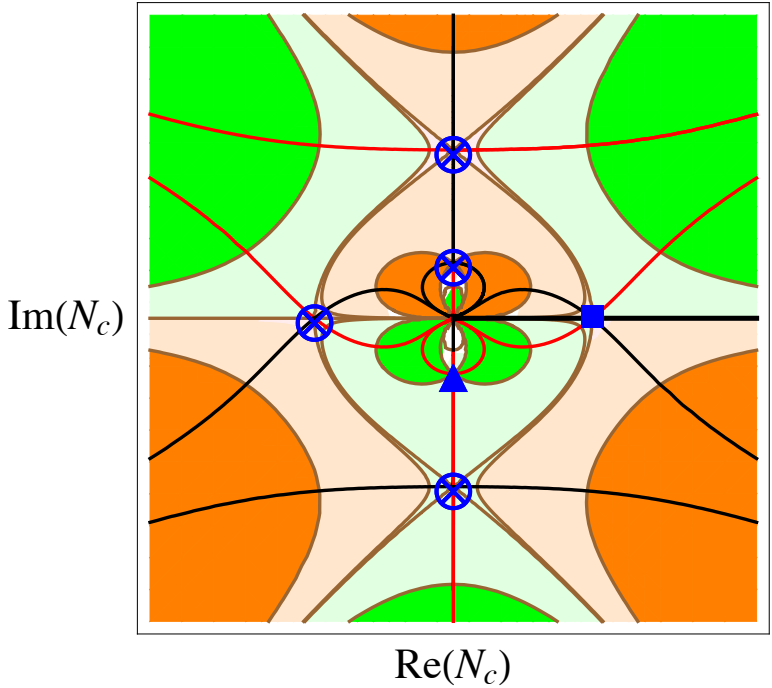}
}
\caption[]{
We consider the case of no-boundary Universe where we choose parameter 
values: $k=1$, $\Lam=3$, $\ep^{\prime\prime}=10^{-2}$ and $\al=2$. 
We take $q_1=3$. 
We plot on $x$-axis real-part of $N_c$ while the $y$-axis is 
imaginary part of $N_c$. 
The red lines correspond to steepest descent lines (thimbles ${\cal J}_\sg$), while 
the thin black lines are steepest ascent lines and denoted by ${\cal K}_\sg$. 
The various saddle points $N_\sg$ are shown in blue. 
The blue cross-circle are \textit{irrelevant} saddle points. 
The saddle point blue-square and blue-triangle are {\it relevant}. 
The steepest ascent line emanating from it can be 
connected to original integration contour. 
$H$ remains constant along the red and black lines, and is equal to 
the value of $H(N_\sg)$. The green region is allowed region 
with $h<h(N_\sg)$ for all values of $\sg$. The orange region 
(forbidden region) has $h>h(N_\sg)$ for all $\sg$. 
The light-green, light-orange and un-colored region has intermediary values. 
The boundary of these region is depicted in brown lines. 
Along these line we have $h = h(N_\sg)$. The original contour of 
integration $(0,\infty^+)$ is shown by thick black line. 
}
\label{fig:nbu_csb}
\end{figure}
%
In figure \ref{fig:nbu_csb} we consider an example of the above scenario.
We plot the set of saddle points along with the collection of flow lines 
associated with each saddle. 
The coloring of graph is done obeying the relation between the values 
of Morse-function at various saddle points. 
The region where $h(N_c) < h(N_1^-)$ is colored green. The region where 
$h(N_c) > h(N_1^+)$ is colored orange. The light-green and light-orange 
region has intermediary values. 
The boundary of these regions is depicted in brown lines. 
The steepest descent lines are shown in 
red while the steepest ascent lines are shown in black. 
The thick black-line depicts the original integration contour. 
The upward flow through original integration contour 
only hits the saddle points $N_0^+$, $N_1^-$ and $N_2^-$.
However, only the former two are {\it relevant} with 
corresponding $h<0$, while $N_2^-$ is {\it irrelevant}
with corresponding $h$ becoming positive for $q_1>q_1^{\rm th}$ (threshold 
value). 

The deformed contour of integration can be chosen such that it passes 
through all the {\it relevant} saddle points (depicted in 
blue-triangle and blue-square), and follow 
closely the Lefschetz-thimbles passing though them.
The saddles depicted by blue-triangle is predominantly imaginary, 
and hence correspond to a predominantly Euclidean geometry,
while the saddle point depicted in blue-square is predominantly 
real and hence correspond to a predominantly Lorentzian 
geometry. 

The deformed contour starts at blue-triangle then circles around 
following the Lefschetz-thimble (red-line) lying in lower-right quadrant. 
Then it approaches origin. Thereafter near the origin it turns back,
hovers around in the green region following the steepest-descent line 
approaching the blue-square. Thereafter it follows the Lefschetz-thimble (red line) connecting 
blue-square in the upper-right quadrant. 
For the two {\it relevant} saddle the corresponding $\dot{q}_0$ at the 
initial boundary can be computed using eq. (\ref{eq:q1inq0p}). 
As the $N_c$ at the two saddle point is different, as a result 
the initial $\dot{q}_0$ is different, indicating that the final 
geometry is a super-position of two different initial configurations.

%
\begin{figure}[h]
\centerline{
\vspace{0pt}
\centering
\includegraphics[width=4.5in]{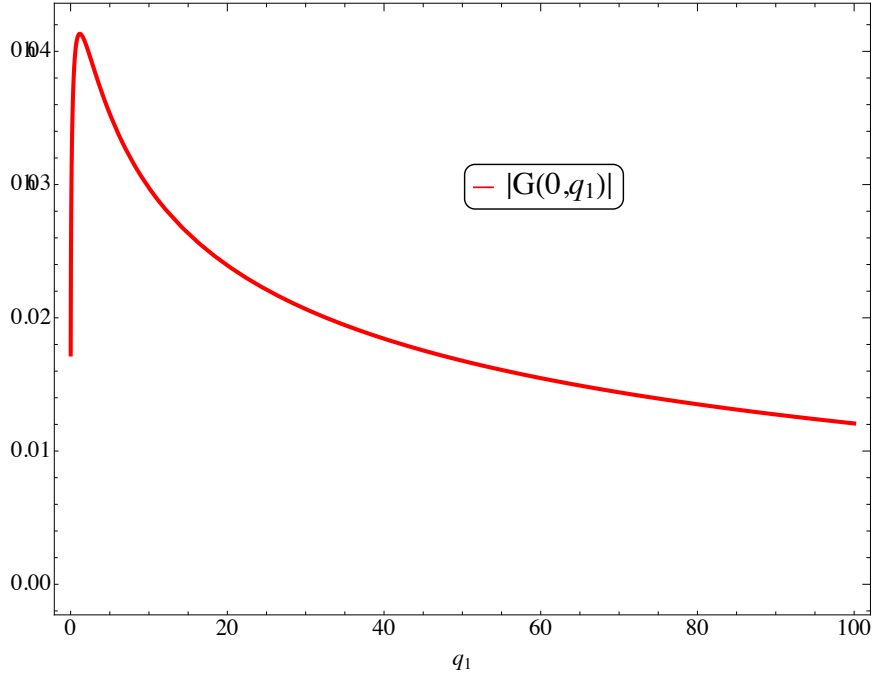}
}
\caption[]{
We consider the case of no-boundary Universe where we choose parameter 
values: $k=1$, $\Lam=3$, $\ep^{\prime\prime}=10^{-2}$ and $\al=2$. Here we plot the 
transition amplitude $G(0, q_1)$ as $q_1$ is varied.  
}
\label{fig:gamp_csb}
\end{figure}
%
We numerically compute the transition 
amplitude in this particular case and plot it in figure \ref{fig:gamp_csb}
as a function of $q_1$. For this situation under consideration we have 
$\ta_{\color{blue}{ \blacktriangle}}=0$ and
$\ta_{\color{blue}{\blacksquare}}=\pi/4$. 
Both saddle-points contribute in exponentially suppressed manner
as $h<0$ for both of them. But this suppression varies with respect to 
$q_1$. At each $q_1$ the weight of blue-square saddle point 
is more than the weight of blue-triangle saddle favouring 
a Lorentzian geometry.

\section{Complex initial momentum}
\label{cmom}

In this section we consider a very simple model of no-boundary proposal 
where we directly fix the initial field derivative $\dot{q}_0$. Certainly, this scenario 
fall in case $(b)$ category as discussed in subsection \ref{mixedBC}. 
To properly motivate the choice of $\dot{q}_0$ we start by considering 
deSitter (dS) geometry which will be solution to bulk equation of motion. 
This means that for $\Lam=3\lam^2>0$ in $d=4$ we have for the 
spacetime dS metric in eq. (\ref{eq:frwmet})
\beq
\label{eq:atp_ds}
N_p = 1 \, , 
\hspace{5mm}
a(t_p) = \frac{1}{\lam} \cosh \left(\lam t_p\right) \, .
\eeq
dS can be embedded in $5$-dimensions where in closed slicing it can 
pictured as hyperboloid having a minimum spatial extent at $t_p=0$. 
The intuition behind the no-boundary proposal is that the geometry 
is rounded off, so as to have no boundary in the beginning of time.
This can be achieved by analytically continuing the original dS metric 
to Euclidean time, starting exactly at the waist of hyperboloid at $t_p=0$. 
This means 
\beq
\label{eq:eucTp}
t_p = \mp i \left(\tau - \frac{\pi}{2\lam} \right) \, , 
\hspace{5mm}
0 \leq \tau \leq \frac{\pi}{2\lam} \, .
\eeq
This means that along the Euclidean section the dS metric transforms 
in to that of a $4$-sphere 
\beq
\label{eq:4sep}
{\rm d}s^2 = {\rm d} \tau^2 + \frac{1}{\lam^2} \sin^2 \left(\lam \tau \right) {\rm d} \OM_3^2 \, .
\eeq
This geometry has no boundary at $\tau=0$ and smoothly closes off. 

It should be emphasised that there are two possibilities of the time rotation
to Euclidean time above, corresponding to the sign appearing in eq. (\ref{eq:eucTp}).
Each of these choices correspond to a different Wick rotation. 
The upper sign correspond to the standard Wick rotation which is also 
used in the flat spacetime QFT. It is also the sign chosen in the work 
of Hartle and Hawking \cite{Hartle:1983ai,Halliwell:1984eu}. For this sign the 
perturbations around the geometry are stable and suppressed. 
The lower sign in eq. (\ref{eq:eucTp}) correspond to Vilenkin's tunneling geometry 
where small perturbation around the geometry are unsuppressed
\cite{Feldbrugge:2017fcc,Halliwell:1989dy}. The process of 
Wick rotation can also be thought of the lapse $N_p$ changing its 
value from $N_p=1$ to $N_p = \mp i$, thereby implying that the 
total time $T_p = \int N_p {\rm d} t_p$ becoming complex valued. 

This can be translated into the language of metric in eq. (\ref{eq:frwmet_changed})
and will thereby imply
\beq
\label{eq:HHlapse_tr}
\sinh\left(\lam t_p \right)= \lam^2 N_{\rm HH} t + i \, ,
\eeq
where $N_{\rm HH}$ will turn out to the saddle-point value of
the lapse integral corresponding to Hartle-Hawking geometry \cite{Hartle:1983ai,Halliwell:1984eu}.
It is given by
\beq
\label{eq:HHsad}
N_{\rm HH} = \frac{\sqrt{\lam^2 q_1 - 1}}{\lam^2} - \frac{i}{\lam^2} \, ,
\eeq
where $q_1 = q(t=1)$. The HH-geometry fall in case $(b)$ of the 
mixed boundary conditions. On comparing it with eq. (\ref{eq:qsol_b}) 
one has $q(t=0)=0$, while the $q(t)$ is given by
\beq
\label{eq:q0dot}
\dot{q}_0 = q_1 - \lam^2 N_{\rm HH}^2 \, ,
\hspace{5mm}
\Rightarrow
\hspace{5mm}
q_{\rm HH}(t) = \lam^2 N_{\rm HH}^2 t^2 + 
\left(q_1 - \lam^2 N_{\rm HH}^2 \right)t \, ,
\eeq
where $0\leq t \leq 1$. From this we can immediately notice 
the complex nature of saddle point value $N_{\rm HH}$ encodes the 
direction of Wick rotation. This can be seen by computing the 
momentum using eq. (\ref{eq:Spot_mom}) at $t=0$
\beq
\label{eq:wickdir}
\left. \frac{\dot{q}_{HH}}{2 N_{\rm HH}} \right|_{t=0} = + i \, .
\eeq
Motivated by the Hartle-Hawking geometry \cite{Hartle:1983ai,Halliwell:1984eu}
where we notice that the initial momentum is complex and appears with 
positive sign resulting in a stable and suppressed behaviour of fluctuations, 
we can consider appling this boundary condition in the case $(b)$ scenario 
that is considered in this paper. More clearly motivated by the works of 
Hartle-Hawking \cite{Hartle:1983ai,Halliwell:1984eu}, 
we choose the following 
mixed boundary condition in the case $(b)$
\beq
\label{eq:mbc_sp_HH}
\dot{q}_0^{(b)} = + 2 i N_c \, ,
\hspace{5mm}
q^{(b)}(t=1) = q_1 \, .
\eeq
Plugging this special condition in eq. (\ref{eq:qsol_b}) and (\ref{eq:SNact_b_mix}) we get
\beq
\label{eq:mix_qb_sp}
q^{(b)}(t) = \lam^2 N_c^2 t^2 + 2 i N_c t 
+ q_1 - 2 i N_c - \lam^2 N_c^2 \, ,
\eeq
and the corresponding action for lapse $N_c$ is given by
\beq
\label{eq:Sbact_sp}
S_{\rm tot}^{\rm HH} = \frac{V_3}{16 \pi G} \biggl[
2\lam^4 N_c^3 + 6 i \lam^2 N_c^2  - 6q_1 \lam^2 N_c
- 2 i \left\{3q_1 + 8\al \right\}
\biggr] \, 
\eeq
respectively. There are few crucial things to note here for this special 
mixed boundary condition: (1) the action for lapse $N_c$ is complex 
(2) the action is no longer singular at $N_c=0$. The former is a direct 
consequence of the imposition of complex initial momentum which 
subsequently leads to complex geometries. A complex action will mean 
that even for real values of lapse $N_c$ there will be a non-zero weighting 
of the corresponding geometrical configuration. 

The later point about the lack of $N_c=0$ singularity can be understood 
by realising that as we are fixing the initial momentum (and not the initial size of geometry). 
As result we are summing over all possible initial $3$-geometry size 
and their transition to $3$-geometry of size $q_1$. This will also 
include a transition from $q_1 \to q_1$. Such a transition can occur 
instantaneously {\it i.e.} with $N_c=0$. This means that there is nothing 
singular happening at $N_c=0$. 

An interesting third observation is that the saddle point equation following 
from action in eq. (\ref{eq:Sbact_sp}) is quadratic in $N_c$. 
\beq
\label{eq:sadpot_sp}
\frac{{\rm d} S_{\rm tot}^{\rm HH}}{{\rm d} N_c} = 0 
\hspace{5mm}
\Rightarrow
\hspace{5mm}
\lam^4 N_c^2 + 2 i \lam^2 N_c  - q_1 \lam^2 = 0 \, .
\eeq
This quadratic equation has only two saddle point solution, unlike the 
scenarios studied in previous section where there were six saddle 
points. In the present case the saddle points are also 
independent of the Gauss-Bonnet coupling $\al$. These
saddle points have a very simple expression 
\beq
\label{eq:sadHH_sp}
N_\pm = \frac{-i \pm \sqrt{q_1 \lam^2 -1}}{\lam^2} \, .
\eeq
It should be noted that $N_+$ is same as the saddle point considered in the 
work of Hartle-Hawking \cite{Hartle:1983ai,Halliwell:1984eu}.

At this point our interest is to compute eq. (\ref{eq:Gab_afterQ}) for the 
case $(b)$ for the boundary condition mentioned in eq. (\ref{eq:mbc_sp_HH}). 
The $N_c$-action is given in eq. (\ref{eq:Sbact_sp}). As the integrand is not 
singular for $N_c=0$, so one can extend the range of the $N_c$-integration 
from $-\infty$ to $\infty$. Then we have
\beq
\label{eq:Gab_HH_sp}
G[\dot{q}_0=2i N_c, q_1]
= \frac{1}{2\sqrt{\pi i}} \int_{-\infty}^\infty {\rm d} N_c \,\, 
\exp \left(\frac{i}{\hbar} S^{\rm HH}_{\rm tot} \right) \, .
\eeq
This can be performed using the Picard-Lefschetz and WKB methods. 
Once the saddle points for the action $S^{\rm HH}_{\rm tot}$ are known, one 
can compute the steepest ascent/descent flow lines corresponding to each of the
saddle point. A saddle point is termed {\it relevant} if the steepest ascent path emanating 
from it hits the original integration contour which is $(-\infty, +\infty)$. 
If the action is real then it implies that 
the {\it relevant} saddle points will have their corresponding Morse-function $h<0$. 
However, in the case when action is complex this obstruction can be evaded.

\begin{figure}[h]
\centerline{
\vspace{0pt}
\centering
\includegraphics[width=4in]{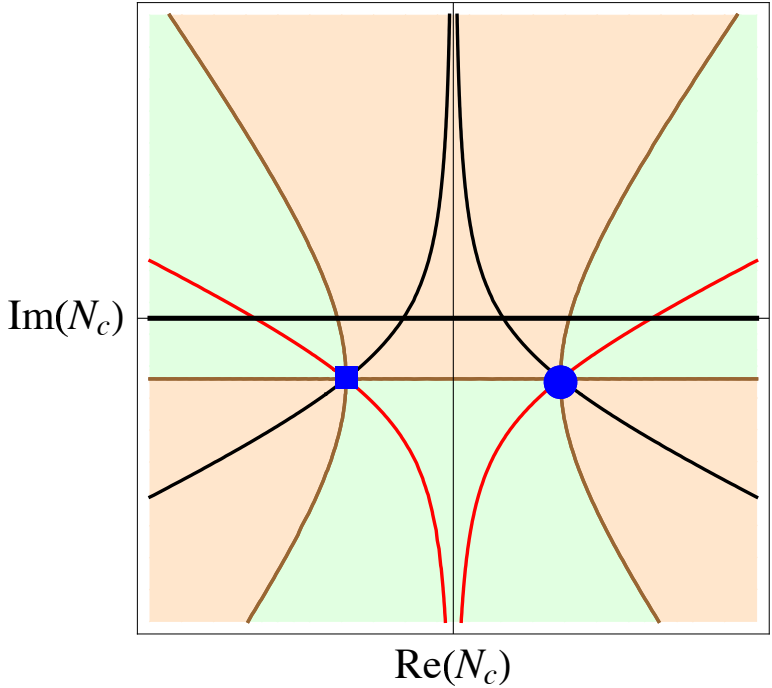}
}
\caption[]{
We consider the case of no-boundary Universe where we impose the mixed 
boundary condition: with Euclidean momentum at $t=0$ and fixed final size at $t=1$. 
The lapse action given in eq. (\ref{eq:sadpot_sp}) is complex. 
We take $\dot{q}_0 = + 2 i N_c$ motivated by work of Hartle-Hawking 
\cite{Hartle:1983ai,Halliwell:1984eu}. For the purpose of this numerical example 
we take $\lam=1$, $\al=2$. We choose final boundary condition 
to be $q_1=3$. We plot on $x$-axis real-part of $N_c$ while the $y$-axis is 
imaginary part of $N_c$. 
The red lines correspond to steepest descent lines (thimbles ${\cal J}_\sg$), while 
the thin black lines are steepest ascent lines and denoted by ${\cal K}_\sg$. 
Both the saddle points are depicted in blue: $N_-$ (blue-square) and 
$N_+$ (blue-circle). Both saddle points are {\it relevant}. The 
steepest ascent lines emanating from both of them intersects the 
original integration contour $(-\infty, +\infty)$ which is shown by
thick-black line. The Morse-function $h$ is same for both saddle points:
$h(N_\pm)>0$. $H$ remains constant along the red and thin-black lines, and is equal to 
the value of $H(N_\sg)$. The light-green region is allowed region 
with $h<h(N_\sg)$ for all values of $\sg$. The light-orange region 
(forbidden region) has $h>h(N_\sg)$ for all $\sg$. 
The boundary of these region is depicted in brown lines. 
Along these line we have $h = h(N_\sg)$.
}
\label{fig:HH_sp}
\end{figure}
%

The analyse the nature of Morse-function at each saddle point we first
compute the on-shell action, which is obtained by plugging the 
saddle point solution given in eq. (\ref{eq:sadHH_sp}) back in the action 
given in eq. (\ref{eq:Sbact_sp}). The on-shell action at the two saddle points is given by,
\beq
\label{eq:Sact_HH_sp}
S^{\rm HH}_\pm = 2 \pi^2 \left[
- i \left(\frac{1}{\lam^2} + 4 \al \right)
\mp \frac{\left(q_1 \lam^2 -1\right)^{3/2}}{\lam^2}
\right] \, .
\eeq
It should be emphasised here that only the imaginary part of the action 
gets correction from the Gauss-Bonnet sector of gravity while the real 
parts remains unaffected and is same as for pure Einstein-Hilbert gravity. 
This immediately implies that for $q_1 > 1/\lam^2$, the Morse-function for the two saddle points is
\beq
\label{eq:Mor_HH}
h(N_\pm) = \frac{2\pi^2}{\hbar} \left(\frac{1}{\lam^2} + 4 \al \right) \, .
\eeq
It is real-positive and independent of $q_1$. However, it receives a correction 
from the Gauss-Bonnet sector of gravity action. By analysing the steepest ascent 
flow lines emanating from both the saddle points it is realised that both of them 
are {\it relevant}. Even though for both of them $h(N_\pm)>0$. 

In figure \ref{fig:HH_sp} we plot the various flow-line, saddle points, forbidden/allowed regions. 
As both the saddle points are relevant, so the corresponding Lefschetz thimbles 
passing through both the saddle points constitute the deformed contour of integration. 
This deformed contour starts at upper-left quadrant, follow the red-line, crosses the 
negative real-axis then goes over to lower-left quadrant and asymptotes to $-i\infty$. The second
part of contour starts $-i\infty$ in lower-right quadrant, follows the red-line, crosses the 
positive real-axis, then goes to the upper-right quadrant following the red-line. 
The Picard-Lefschetz theory then gives the transition amplitude 
in the saddle point approximation as
\bea
\label{eq:GHH_sp_exp}
&&
G[\dot{q}_0=2i N_c, q_1]
= \frac{1}{2\sqrt{\pi i}} \left[ \exp\left(\frac{i S^{\rm HH}_{\rm tot}(N_-)}{\hbar}\right) 
+\exp\left(\frac{i S^{\rm HH}_{\rm tot}(N_+)}{\hbar}\right) \right]
\notag \\
&&
= \frac{e^{-i\pi/4}}{\sqrt{\pi}}\exp\left[\frac{2\pi^2}{\hbar} \left(\frac{1}{\lam^2} + 4 \al \right)\right]
\cos \left[\frac{2\pi^2 \lam}{\hbar} \left(q_1 - \frac{1}{\lam^2}\right)^{3/2} \right] \, .
\eea
We notice that we get a non-perturbative correction from the Gauss-Bonnet sector 
of gravity to the Hartle-Hawking wave-function from a Lorentzian path-integral. 
This transitional amplitude is fully non-perturbative and incorporates the 
non-trivial features coming from the Gauss-Bonnet coupling.

\section{Conclusion}
\label{conc}

In this paper we study the path-integral of gravitational theory where the gravitational 
dynamics is governed by Einstein-Hilbert gravity with an addition of 
Gauss-Bonnet gravity. We study this setup in four spacetime dimensions 
directly in Lorentzian signature. In four spacetime dimensions the 
Gauss-Bonnet sector of gravity action is also topological and doesn't contribute in the 
bulk dynamics. However it has a crucial role to play at boundaries. 
Depending on the nature of boundary conditions the Gauss-Bonnet modifications 
will affect the study of path-integral. This paper aims to investigate these issues 
by considering the gravitational path-integral in a reduced setup of 
mini-superspace approximation. 

We start by considering the mini-superspace action of the theory and vary
it with respect to field variables to study the dynamical equation of motion and 
the nature of boundary terms. To have a consistent boundary value problem 
one has to incorporate additional terms at the boundary. We notice that with
Neuman boundary condition one ends up with inconsistencies in fixing the 
free parameters in the solution to the equation of motion. With Dirichlet boundary 
conditions on other hand it is seen that no non-trivial effects arise from the Gauss-Bonnet sector.
However, in the case of mixed boundary conditions (where one specifies $q(t)$ at one 
end point and its derivative $\dot{q}(t)$ at another end point) we notice that the Gauss-Bonnet 
sector starts to play a non-trivial role. Although the solution to the equation of motion for 
$q(t)$ doesn't get contribution from the Gauss-Bonnet sector, however, the action 
for the lapse $N_c$ gets non-trivial addition due the non-vanishing boundary
terms. Such non-trivialities arising from the Gauss-Bonnet sector later leads 
to richer features while evaluating the integration over lapse $N_c$
in eq. (\ref{eq:Gamp}).

The paper aims to study the transition amplitude from one 
$3$-geometry to another and investigate the circumstances under which
the Gauss-Bonnet sector starts to affects this amplitude in a non-trivial manner. 
Such a transition amplitude is dictated by a path-integral over 
$q(t)$ and a contour integration over lapse $N_c$. The path-integral 
over $q(t)$ can be performed exactly as the Gauss-Bonnet part 
controls only the boundary, while the bulk remains unaffected. The 
path-integral over $q(t)$ is governed entirely by the Einstein-Hilbert part of 
gravity action. Once the integral over $q(t)$ respecting the boundary conditions 
is performed, we are left with an contour integration over lapse $N_c$ 
given in eq. (\ref{eq:Gab_afterQ}), with the integrand containing 
non-trivial features coming from the Gauss-Bonnet sector. 

We analyse this contour integration by lifting lapse $N_c$ to complex plane
and making use of Picard-Lefschetz theory to investigate the nature of
integrand. We find the saddle points of the $N_c$-action 
and realise that they occur in three pairs. This is a new 
feature of the Gauss-Bonnet gravity which is absent in the case of pure 
Einstein-Hilbert gravity having only two (or less) pairs of saddle points.
In the mixed boundary conditions case the Gauss-Bonnet sector 
contributes non-trivially and give rise to additional saddle points
in the complex $N_c$ plane. The three pairs of saddle points 
follow from the cubic saddle-point equation in $N_c^2$. 
Moreover, if the cubic polynomial equation has real coefficients then 
the nature of saddle points can be determined by analysing the 
discriminant $\D$ of the cubic equation, which 
in turn depend on parameter values and boundary conditions. 

As an application of this we considered an example of no-boundary 
Universe, and analyse the transition amplitude in this setup. 
In this situation the initial $q_0=0$. This has consequences: in 
case $(a)$ $\dot{q}_1^{(a)}$ and $q_1^{(a)}$ are related, 
while in case $(b)$ $\dot{q}_0^{(b)}$ and $q_1^{(b)}$ are related.
In either case we have multiple {\it relevant} saddle points, 
so this implies that for real-positive $q_1$ in the case $(a)$ we will 
have multiple possibilities for $\dot{q}_1^{(a)}$. This is 
contradictory to our original boundary condition 
requirement where $\dot{q}_1^{(a)}$ is supposed to be fixed at the final boundary.
Case $(a)$ is therefore ruled out. Such a contradiction doesn't happen in case $(b)$. 
In case $(b)$ for a fixed real-positive $q_1^{(b)}$ there are 
multiple value for $\dot{q}_0^{(b)}$ for the corresponding 
{\it relevant} saddle-points. This is acceptable as the final 
geometry can be seen as arising from the superposition of 
multiple allowed initial configurations. 

In case $(b)$ by making use of the relation in eq. (\ref{eq:q1inq0p})
one can obtain the action for lapse $N_c$ entirely in terms of 
$q_1$. As $q_1$ is positive, so the action for lapse is entirely real. 
The saddle point equation is cubic in $N_c^2$ with real coefficients. 
We realise that for positive $k$, $\Lam$, and $\al$ there
are three distinct real roots for $N_c^2$: one positive and two negative. 
This implies that one saddle point is always real-positive while 
its `twin' is real-negative. There are four saddle points lying 
on imaginary axis: two on positive imaginary axis while their 
conjugate twins on negative imaginary axis. There are total of 
six saddle-points, which is new compared to the pure Einstein-Hilbert gravity.
Attached to each saddle-point there are two Lefschetz thimbles 
and two steepest ascent lines. Only three of the saddle point 
can be reached by flowing downward along the steepest ascent lines 
starting from the original integration contour $(0^+, \infty)$. 
Out of the three only two have their corresponding Morse-function $h<0$,
making them {\it relevant}. One of the {\it relevant} saddle-point lies on negative imaginary 
axis while other lies on positive real-axis. The deformed contour of 
integration passes through these {\it relevant} saddles following the
Lefschetz-thimbles. The deformed contour therefore incorporate 
contributions from both saddles resulting in interference. 
The full amplitude is a superposition of the contribution coming from 
two configurations with the more weight associated to Lorentzian saddle 
compared to Euclidean saddle. 

We consider another special case of no-boundary 
proposal with a complex initial momentum. Here we are inspired by the 
past works of Hartle-Hawking \cite{Hartle:1983ai,Halliwell:1984eu},
where the authors noticed that a particular choice of Wick-rotation 
leads to stable and suppressed perturbations. This choice of Wick-rotation 
eventually implies that the initial momentum in the cosmic 
evolution was complex. Inspired by their work we choose 
a special initial boundary condition to be $\dot{q}_0^{(b)} = + 2 i N_c$.
This particular scenario falls in the category of case $(b)$ of mixed 
boundary conditions. We notice that in particular in this situation 
the lapse $N_c$ action is complex, and that the action is 
non-singular at $N_c=0$. Moreover, in this case we have only 
two saddle-points and both are {\it relevant}. We compute 
the transition amplitude from the initial to final configuration 
and obtain an analogue of Hartle-Hawking wave-function 
having a non-trivial and non-perturbative correction from the Gauss-Bonnet
sector of gravity theory. 

Certainly, more work needs to be done in this direction as many things 
are still unexplored.  
In the study in section \ref{nbu} we haven't directly fixed the initial line $\dot{q}_0$
in the case $(b)$, rather sort of derived it by imposing condition that 
$q_0^{(b)}=0$ and $q_1^{(b)}>0$. This two requirements eventually 
leads to two {\it relevant} saddle points. For each of these saddle point 
there is a corresponding fixed $\dot{q}_0^{(b)}$. This is hinting at fact 
that the final geometry of Universe is arising due to superposition 
of the two very different initial configurations. Perhaps there are 
two different copies of Universe initially whose evolution and 
interference resulted in the final geometry of the Universe. 
Were these two Universe entangled in past and overtime this 
entanglement grew stronger resulting in current Universe? 
This is hard to answer in present manuscript. 

Another crucial thing missing in this paper is an analysis about the 
behaviour of fluctuations, which are important to understand the 
stability of Universe. In past works on no-boundary Universe 
it was noticed that such models are unstable to fluctuations \cite{Feldbrugge:2017fcc}. 
This is a worrisome feature which if it exists make the 
model less reliable. Past attempts to overcome these issues 
involved imposing different types of boundary conditions 
for background and for fluctuations \cite{DiTucci:2019bui,DiTucci:2019dji}.
It is worth asking this same question 
in the case of the Gauss-Bonnet gravity too. Does the Gauss-Bonnet 
modifications leads to a more stable behaviour of fluctuations? 
If not then what kind of boundary conditions should be imposed for the 
fluctuations? Moreover, in the case of HH-model the choice of Wick-rotation 
leads to stable and suppressed behaviour of fluctuations 
\cite{Hartle:1983ai,Halliwell:1984eu}. Currently it is not clear whether 
these fluctuations will remain suppressed when 
non-perturbative corrections from Gauss-Bonnet gravity are incorporated. 
We plan to address this in our future work.

\bigskip
\centerline{\bf Acknowledgements} 

I will like to thank Jean-Luc Lehners, Alok Laddha, Nirmalya Kajuri and Avinash Raju for useful discussions. 
GN is supported by ``Zhuoyue" (distinguished) Fellowship (ZYBH2018-03).

%


\end{document}